\documentclass[useAMS, usenatbib]{mn2e}
\usepackage{amsmath}
\usepackage{amssymb}
\usepackage{graphicx}  

\DeclareMathOperator{\sech}{sech}   
\def\msun{\,M_\odot}
\def\pc{\,{\rm pc}}
\def\kpc{\,{\rm kpc}}
\def\kms{\,{\rm km\,s^{-1}}}
\def\vlos{v_{\rm los}}
\def\avlos{\left|\vlos\right|}
\title[High velocity bar stars]{Origin of the high $\vlos$ feature in the Galactic bar}
\author[M. Aumer and R. Sch\"onrich]
{Michael Aumer \thanks{E-mail:Michael.Aumer@physics.ox.ac.uk (MA)} and Ralph Sch\"onrich\\
Rudolf Peierls Centre for Theoretical Physics, 1 Keble Road, Oxford, OX1 3RP, UK}

\begin{document}

\date{Accepted 2015 September 28.  Received 2015 September 7; in original form 2015 June 19}

\pagerange{\pageref{firstpage}--\pageref{lastpage}} \pubyear{2015}

\maketitle
     
\label{firstpage}

\begin{abstract}

We analyse a controlled $N$-body + smoothed particle hydrodynamics simulation of a growing disc galaxy within a non-growing, live dark
halo. The disc is continuously fed with gas and star particles on near-circular orbits and develops a bar comparable in size 
to the one of the Milky Way (MW). We extract line of sight velocity $\vlos$ distributions from the model and compare it to data
recently obtained from the APOGEE survey which show distinct high velocity features around $\vlos\sim 200\kms$.
With an APOGEE like selection function, but without any scaling nor adjustment, we find $\vlos$ distributions very similar to those in APOGEE.
The stars that make up the high $\vlos$ features at positive longitudes $l$ are preferentially young bar stars (age $\tau\lesssim 2-3\;{\rm Gyr}$)
which move away from us along the rear side of the bar. At negative $l$, we find the corresponding low $\vlos$ feature from stars moving towards us.
At $l>10$ degrees the highest $\vlos$ stars are a mixture of bar and background disc stars which complicates the interpretation of observations.
The main peak in $\vlos$ is dominated by fore- and background stars. At a given time, $\sim 40-50$ per cent of high $\vlos$ stars occupy 
$x_1$ like orbits, but a significant fraction are on more complex orbits. The observed feature is likely due to a population 
of dynamically cool, young stars formed from gas just outside the bar and subsequently captured by the growing bar.
The high $\vlos$ features disappear at high latitudes $\left|b\right|\gtrsim2$ degrees which explains the non-detection of such features in other surveys.
 
\end{abstract}


\section{Introduction}

The central region of our Galaxy is dominated by a bar \citep{blitz,binney91}, which consists of an X-shaped
box/peanut bulge at $R<2\kpc$ \citep{wegg13} and a vertically thin part, the {\it long bar}, extending 
to $R\sim 4-5\kpc$ \citep{wegg15}. Recently, the Apache Point Observatory Galactic Evolution Experiment (APOGEE)
 survey \citep{apogee} has revealed that the line of sight velocity $\vlos$ distributions
of $\sim 4\,700$ K/M giant stars towards the Galactic centre at Galactic coordinates $l=4-14$ degrees and $\left|b\right|\le 2$ degrees
exhibit distinct shoulders at high velocities $\vlos\sim 200\kms$ \citep{nidever} .
This result was followed up by other surveys, which yielded differing results: On the one hand, \citet{babusiaux} analysed 
the $\vlos$ distribution of $\sim 400$ bar red clump stars at $l=(-6)-10$ degrees and $\left|b\right|\le 1$ degrees
and found hints of a distinct high $\vlos$ component at $\vlos>200\kms$ at positive $l$ and 
additionally of a distinct low $\vlos$ component at $\vlos<-200\kms$ at negative $l$.
On the other hand, \citet{zoccali} failed to detect distinct high or low $\vlos$ components in a survey
of $\sim 5\,000$ red clump stars in the central Galaxy. However, they were probing
different sight-lines, most of them at $\left|b\right|>2$ degrees and thus further away from the plane
than the other authors, or pointing more closely towards the Galactic centre at $l=0$.

Bars are rotating triaxial structures which are expected to be supported mostly by stars on prograde
orbits parented by the $x_1$ family of closed long-axis orbits \citep{contopoulos}. These orbits have proven
essential in explaining the $(l,v)$ diagram of gas towards the Galactic centre \citep{binney91} and are thus expected
to be an important ingredient for explaining any observation of the kinematics of bar stars.
\citet{molloy} have recently shown that they could recover a distinct high $\vlos$ peak in an $N$-body simulation
of a barred galaxy if they analysed only 2:1 resonant orbits elongated along the long bar axis, i.e.
$x_1$ orbits. They did not recover the feature when taking into account all stars. Previously, 
\citet{nidever} and \citet{li} had failed to recover the observational feature in $N$-body simulations
of Milky Way (MW) like galaxies. This led \citet{li} to conclude that the high $\vlos$ peak is either
a spurious feature due to low numbers of stars or not connected to stars on bar orbits.

Here we analyse a controlled simulation of a growing MW-like disc galaxy with respect to $\vlos$ distributions
towards the barred galactic centre. The model galaxy features a live dark halo and is continuously fed with stellar and
gas particles on near circular orbits. This type of simulation was introduced by \citet{selcar} for a 2D disc but has scarcely
been used since (e.g. \citealp{berrier} for a specific problem). The disc develops a bar early on and 
becomes comparable in structure to the MW at later stages, which allows us to create $\vlos$ distributions
and compare them to the APOGEE data.

With the help of this simulation, we explore the origin of the high $\vlos$ structures and fit the detections 
and non-detections into a consistent picture. We find that these features are associated with a kinematically cool
population of stars running alongside the bar and likely recently captured from the Galactic disc as a consequence of bar growth.

Our paper is organised as follows:
Basic information about the data and simulation are found in Section 2 (survey data, selection function) and Section 3 (simulation setup).
The general evolution of disc and bar in our simulation is discussed in Section 4.
Detailed kinematics are analysed in Section 5 with focus on the high $\vlos$ peak.
In Section 6 we apply the APOGEE selection function and directly compare the model to the data. We also explore some variation of assumed parameters, e.g. latitude $b$ and bar angle $\phi$.
In Section 7, we use the successfully identified high $\vlos$ peak for a detailed study of the contributing orbits.
Finally, we conclude in Section 8.

\begin{figure}
\centering
\vspace{-0.cm}
\includegraphics[width=8cm]{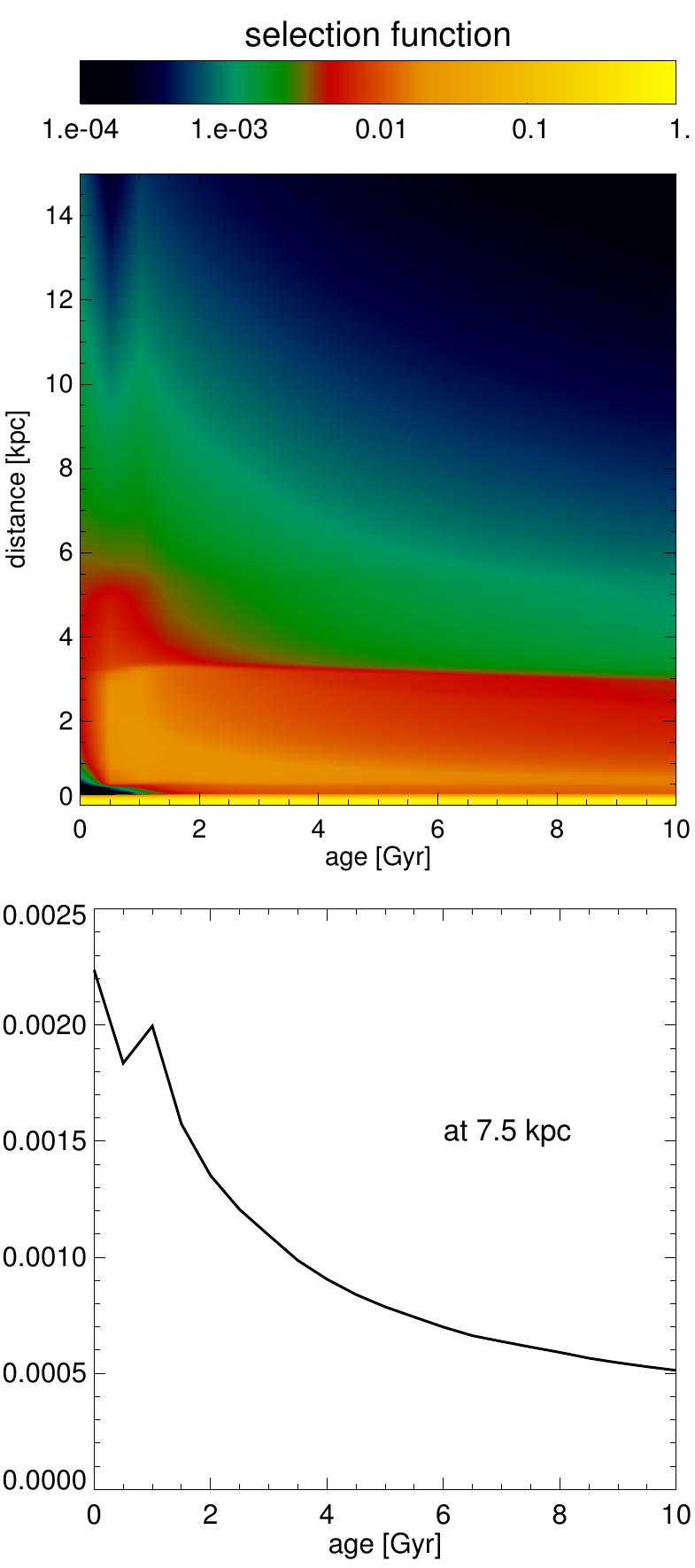}
\caption
{The Selection function in distance $s$ and age $\tau$ at a fixed metallicity of $Z = 0.0217$. 
{\it Top panel}: Contour plot as a function of $s$ and $\tau$.
{\it Bottom panel}: The age dependence at a typical bar distance of $s=7.5\kpc$.
Note the strong bias towards stars with ages $< 2 $ Gyr at larger distances.
}
\label{selfu}
\end{figure}

\section{Observations and selection function}

For kinematic data from the central bar/bulge region of the MW we use data from the Apache Point Observatory Galactic Evolution Experiment (APOGEE, \citealp{majewski}), 
an infrared, high-resolution survey, which is part of the Sloan Digital Sky Survey III \citep[SDSS][]{SDSSIII}. The fibre-spectrograph allows for $300$ spectra per plate 
covering a spectral range of $1.51-1.7 \mu m$ in the $H$ band. As an infrared survey, APOGEE has a unique capability to obtain information from stars in the high-extinction 
regions within the Galactic plane and towards the central region of the MW. We use the APOGEE data publicly available via the CAS for SDSS DR10 \citep{ahn}.
Throughout this work, we correct all observed stellar velocities from APOGEE to the Galactic standard of rest, using $(U,V,W)_{\rm Sun} = (14.0, 250.0, 6.0)$, in concordance with
\citet{ralph} and \citet{paul}.

Another advantage of the survey is the benign selection function described in \cite{apogeetarget}. Stars in the target magnitude range with dereddened colours $J-K_s > 0.5\; {\rm mag}$ are targeted 
evenly. In the case of the central Galaxy fields used within this work, stars were targeted for $6 < H < 11$. To model the 
selection function we employ the population synthesis machinery from the \cite{SB} model. We apply a standard Salpeter IMF with an exponent of $\gamma = -2.35$, neglecting 
binarity of stars. The population synthesis uses a dense grid of B.A.S.T.I. stellar models (\citealp{Piet04, Piet09}, for their colour calibrations see \citealp{Bedin05}).
At each distance $s$ and age $\tau$ we count the number of stars per mass of the initial population that would end up within the colour and magnitude selection of APOGEE. The result 
is shown in Figure \ref{selfu} for a fixed metallicity of $Z=0.0217$. 
The use of a single, slightly super-solar metallicity is a simplification, as the inner MW has a relatively wide metallicity distribution around this value \citep{hill, bensby}.
However, all relevant metallicities show a similar preference for the detection of young stars. As the detailed age-metallicity distributions for the relevant sightlines are 
not well known and the detailed connection between model galaxy and MW in terms of structure and formation history is unclear, we choose to keep the model simple and use one typical metallicity.
The band of high selection probabilities at very small distances derives from main-sequence dwarf stars. The stellar
 models are not fully reliable in this region for such cool dwarfs. However, this foreground is strongly suppressed by the geometric factor ($s^2$) of the pencil beam survey.
The number of stars currently in the giant stages of interest to us declines strongly (by roughly one order of magnitude) with age, implying a strong bias in favour of young stars within the APOGEE survey. 
We apply this selection function to our $N$-body simulation by weighting each particle with its respective selection probability.
APOGEE uses dereddened magnitudes for targeting, however there are residual effects from reddening uncertainties, which we neglect here.

\section{Simulations}

In this paper we present a controlled collisionless+hydrodynamic simulation of a growing galactic disc
embedded in a non-growing dark matter halo. The simulations were all carried out with the Tree Smoothed Particle Hydrodynamics (TreeSPH)
code GADGET-3, last described in \citet{gadget2}. The applied gravitational softening lengths are $\epsilon_{\rm disc}=30\pc$ for stellar and gaseous disc particles
and $\epsilon_{\rm DM}=100\pc$ for halo particles. We use an opening angle for the tree code of $\theta=0.5$.
An adaptive time-stepping scheme which assigns time steps in bins $j$ according to 
$\Delta t_i = \tau/2^j < \sqrt{2\eta \epsilon_i /  \left|{\bf{a}}_i\right|}$
to individual particles is applied. Here $\eta=0.02$ is an accuracy parameter and ${\bf{a}}_i$ is the gravitational acceleration 
at the position of particle $i$ with softening $\epsilon_i$. The base timestep $\tau$ is appropriately chosen 
by the code and a maximum of 10 Myr is allowed for timesteps. The hydrodynamical timestep is based on a Courant-like condition
$\Delta t_{i,{\rm hyd}} \propto \kappa h_i/c_s$, where $\kappa=0.15$ is the Courant parameter, $h_i$ is the smoothing length
and $c_s$ is the soundspeed. We apply an isothermal equation of state $P=\rho c_s^2$ with $c_s=20\kms$ and
use $N_{\rm SPH}=48$ neighbours for the smoothing kernel. For further code details we refer to \citet{gadget2}.

\subsection{Initial Conditions}

To generate initial conditions (ICs) for our numerical experiments we use the publicly available GalIC code \citep{yurin},
which produces near-equilibrium ICs of multi-component collisionless systems with given density distributions using
an iterative approach.

The initial system consists of a dark halo with a mass of $M_{\rm{DM}}=0.995\times10^{12}\msun$ represented by $N_{\rm{DM}}=1\times10^{6}$ particles
and an embedded thin galactic disc with a mass of $M_{\rm{disc,i}}=0.5\times10^{10}\msun$ represented by $N_{\rm{disc,i}}=5\times10^{5}$ particles, so that the
disc particle mass is $m_{\rm disc}=1\times10^4\msun$.

The dark halo has a Hernquist profile with 
\begin{equation}
\rho_{\rm{DM}}(r)={{M_{\rm{DM}}}\over{2\pi}} {{a}\over{r\left(r+a\right)^3}}.
\end{equation}
The inner profile is adjusted so that it is similar to an NFW profile with concentration $c=9$ and virial radius $R_{200}=162.6\kpc$.
This leads to a scale radius $a=30.25\kpc$. The halo initially has a spherical density profile and radially isotropic kinematics,
i.e. equal velocity dispersion in the principal directions, $\sigma_{r}=\sigma_{\phi}=\sigma_{\theta}$ and consequently a value of the
anisotropy parameter of $\beta=1-\left(\sigma_{\theta}^2+\sigma_{\phi}^2\right) / \left(2 \sigma_{r}^2\right)=0$.

The disc is set up with a mass profile
\begin{equation}
{\rho_{\rm{disc,i}}(R,z)} = {{M_{\rm{disc,i}}}\over{4\pi h_z{h_R}^2}} {\sech^2 \left({z}\over{h_z}\right)} {\exp\left(-{R}\over{h_R}\right)},
\end{equation}
with an exponential scalelength of $h_R=1.5\;\kpc$ and an isothermal vertical profile with a radially constant scaleheight of $h_z=107\;\pc$.
The vertical velocity dispersion $\sigma_z$ thus declines with radius. We assume $\left({\sigma_z}/{\sigma_R}\right)^2=0.5$,
so that Toomre's $Q$ shows a minimum value of $Q_{\rm min}=1.15$ at $R\sim h_R$.

For further information regarding the ICs we refer to Aumer et al. (in prep.).
We note that here the halo is rather poorly resolved with $N_{\rm{DM}}=1\times10^{6}$ and a rather short softening length ($\epsilon_{\rm halo}=100\pc$) is applied.
The low $N_{\rm{DM}}$ is due to limited computational resources and a preference to use them to resolve the disc. As expected (e.g. \citealp{dubinski}),
purely collisionless (no SPH) tests with a higher $N_{\rm{DM}}=5\times10^{6}$ show reduced disc heating and less bar slowdown.
Yet the disc heating for the simulation studied here is comparable to the MW, with similar velocity dispersions for 10 Gyr old solar neighbourhood stars, since 
heating by massive halo particles compensates the lack of heating by Giant Molecular Clouds or other sources such as satellite galaxies.
As far as bar slowdown due to angular momentum transfer to the halo is concerned, the radius and rotation frequency of bars differ 
by less than $\sim$ 20 per cent at late times (6-10 Gyr), the time relevant for this work.

\begin{figure*}
\centering
\vspace{-0.cm}
\includegraphics[width=18cm]{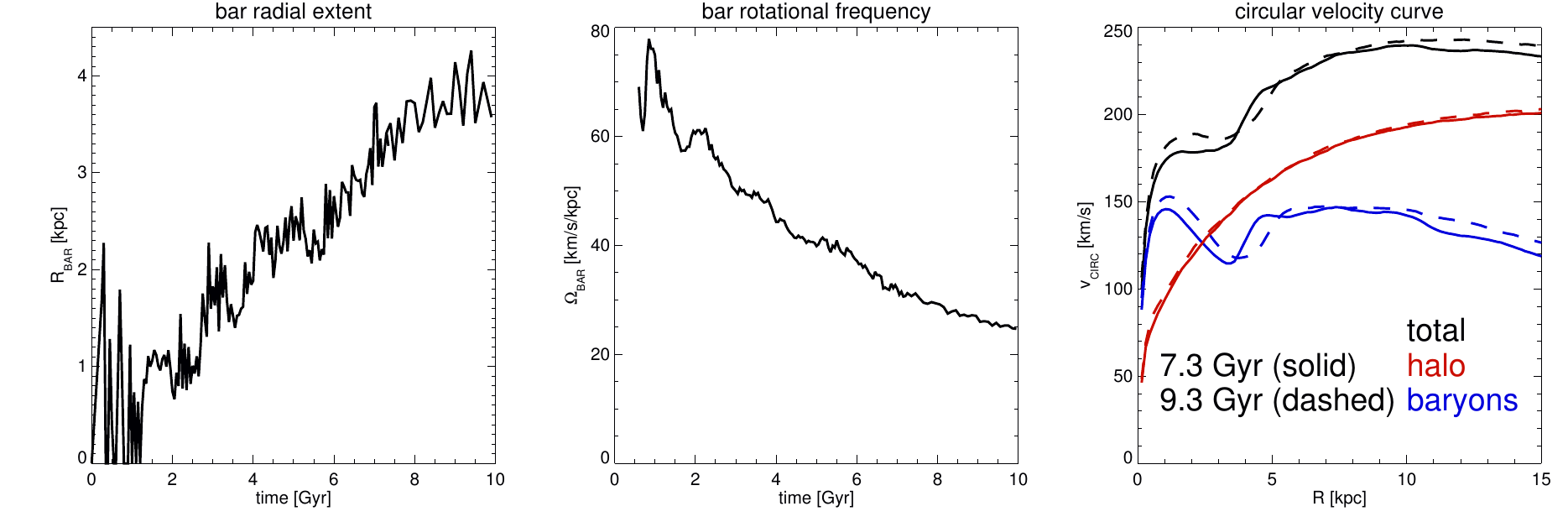}\\
\caption
{The evolution of bar parameters with time:
{\it Left}: The bar radial extent $R_{\rm bar}$, defined as the radius where the logarithm of the $m=2$ Fourier amplitude $\ln(A_2)$ drops below -1.5.
{\it Middle}: The rotational frequency of the bar $\Omega_{\rm bar}$.
{\it Right}: The azimuthally averaged circular velocity curve $v_{\rm circ}(R)$ in the plane of the model galaxy at 7.3 Gyr (solid) and 9.3 Gyr (dashed). The contributions from
                   the dark halo are shown in red and the contributions from the baryons are shown in blue.
}
\label{bar-evo}
\end{figure*}

\subsection{Feeding the disc}

To model the continuous growth of galactic discs via star formation and gas accretion we, over a timespan of 10 Gyr, add new particles with $m_{\rm disc}=1\times10^4\msun$ to the existing
disc every 5 Myr. We assume a mass growth rate history given by
\begin{equation}
  \dot{{\rm M}} (t)= \dot{{\rm M}}_0 \times \exp\left(-{{t}\over{t_{\rm feed} }}\right),
\end{equation}
with an exponential decay timescale $t_{\rm feed}=8\,{\rm Gyr}$, which is motivated by the findings of \citet{ab09},
and a normalisation $\dot{\rm M}_0=8\msun/{\rm yr}$, so that a total of $\sim4.5\times10^{10}\msun$ are fed over 10 Gyr.
The added disc mass is radially distributed as $\rho(R)\propto \exp\left(-R/h_R(t)\right)$, with an exponential scale-length
that increases with time as 
\begin{equation}
h_R(t) = \left(1.5 + {{\left(h_{R, {\rm final}}-1.5\right)}\over{\sqrt{10}}} \sqrt{{t}\over{1\,\rm Gyr}}\right) \kpc.
\end{equation}
This increase from $h_R=1.5\kpc$ to a scale-length at 10 Gyr $h_{R, {\rm final}}=4.3\kpc$ is motivated by the findings of \citet{bovyrix}.
The particles are randomly scattered over azimuth $\phi$ and placed at $z=0$.
The coordinate system is regularly updated to be centred on the centre of mass of the system.

The particles are assigned near circular orbits. We determine the new particles' rotational velocities as $v_{\phi,i}=\sqrt{{a_{R,i}} R_i}$ for a particle $i$
at cylindrical radius $R_i$ and the component of the gravitational acceleration pointing towards the centre, ${a_{R,i}}$. For stellar particles, 
we add random velocity components in all three directions $\phi$, $R$ and $z$, drawn from Gaussian distributions with $\sigma=6\kms$.

After a disc goes bar unstable, there are no stable circular orbits in the bar region. So we impose an inner cutoff radius $R_{\rm cut}$, within which
no particles are introduced. We use the radius where the logarithm of the $m=2$ Fourier Amplitude $\ln(A_2)$ drops below $-1.5$ as an inner cutoff radius. 

\subsection{The gas component}

In disc galaxies, bars slow down and grow in size due to the transfer of angular momentum to the dark halo \citep{debattista}. This loss is to some extent balanced 
by the transfer of angular momentum from the gas to the bar \citep{berentzen}. Therefore, it is desirable to include a gas component in barred
disc galaxy models. We do not wish to introduce a full model containing the accretion of gas onto the galaxy, its cooling and the subsequent
star formation and feedback processes (see e.g. \citealp{roskar} for an isolated galaxy model), as uncertainties regarding the 
modelling of the physical processes are high (see e.g. \citealp{fraternali,nelson}).

To keep the model simple, we add a constant fraction of gas to our galactic disc and model it with an isothermal equation of state $P=\rho c_s^2$ with $c_s=20\kms$.
We simulate the growth of the disc by continuously adding stellar and gas particles. The initial gas disc is created by turning 5 per cent randomly chosen stellar
particles into SPH particles. This mild disturbance of the ICs is unimportant compared to the rapid onset of disc instability in the growing galaxy.
During disc growth we aim at a gas mass fraction in the disc of 10 per cent as observed in modern day MW like disc galaxies.
We achieve this by adding 10 per cent of new particles as gas when the mass fraction of gas in the galaxy is above 10 per cent and 20 per cent of new particles
as gas whenever it is lower.

Computationally, high gas densities in the central galaxy are very expensive. Therefore we include a number of measures to limit the central density:
i) Prior to the bar forming and the inner cutoff on particle insertion coming into play, we create no gas particles at $R<1\kpc$. 
ii) We include a model for star formation in the very central galaxy. We turn SPH particles with hydrogen number densities
$n>n_{\rm th}=10 \;{\rm cm}^{-3}$ and specific angular momentum $j<j_{\rm th}=100\kms\kpc$ into star particles with a probability $p=\eta t_{\rm dyn}/\Delta t_i$,
where $\eta=0.1$ is an efficiency parameter, $t_{\rm dyn}=1/\sqrt{4\pi G \rho}$ is the local dynamical timescale and $\Delta t_i$ is the timestep of
the particle. To mimic the effect of a central galactic outflow \citep{bland} we only turn one out of five particles into a star and remove
the other 4 out of five from the simulation. Note that only gas particles in the inner bar region are affected by these prescriptions.

\section{Structure of the barred disc}

In this section we discuss the structure of our model of a barred disc galaxy. Note that we are comparing a specific - and in no way fitted -
model to a specific galaxy, the Milky Way. Though our model should not provide a perfect match to our Galaxy, we expect the model to qualitatively share important 
characteristics.
Note that due to computational limitations we currently have only one model of this type including hydrodynamics available.

\begin{figure*}
\centering
\vspace{-0.cm}
\includegraphics[width=18cm]{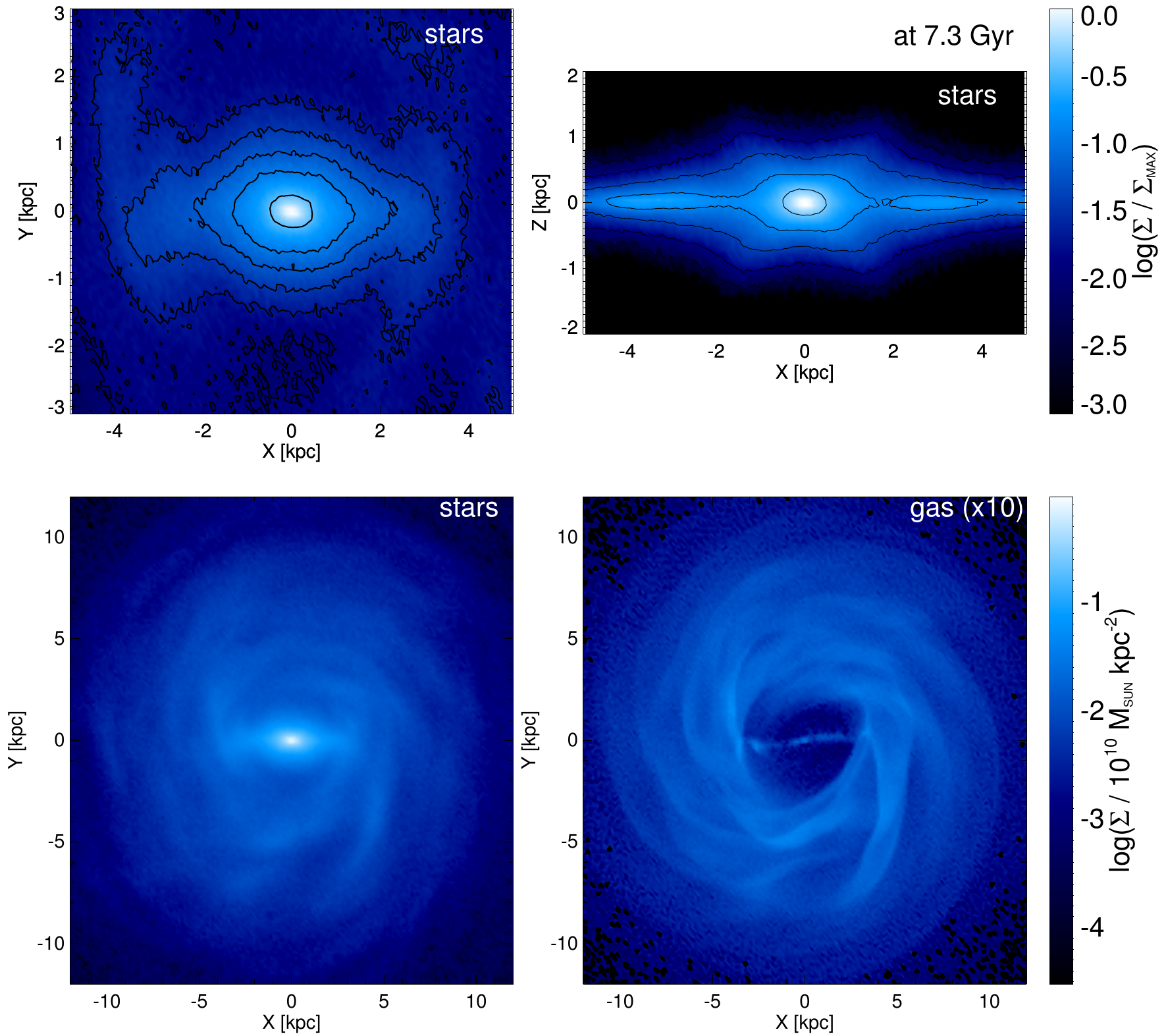}\\
\caption
{Stellar surface density $\Sigma$ projections of our galaxy model at 7.3 Gyr. 
{\it Top Left}: A Face-On view of the bar region.
{\it Top Right}: An Edge-On view of the bar region. Only stars within 2 kpc of the $x-z$ plane are considered.
{\it Bottom Right}: A Face-On view of the stellar disc.
{\it Bottom Right}: A Face-On view of the gas disc. The gas surface density was enhanced by a factor 10.
}
\label{images}
\end{figure*}

In Figure \ref{bar-evo} we present the evolution of bar parameters in our simulation. We determine a bar radial extent $R_{\rm bar}$
by first calculating the $m=2$ Fourier amplitude profile
\begin{equation}
  A_2(R)={{1}\over{N(R)}} \sum\limits_{j=1}^{N(R)}e^{2i\phi_j},
\end{equation}
where N(R) is the number of particles in the radial bin centred on $R$ and we only take into account stellar particles.
Then we define $R_{\rm bar}$ as the radius where $\ln A_2(R)$ drops below -1.5 or as zero if it never reaches the value in the central area.

In the left panel of Figure \ref{bar-evo} we see that the bar forms in the first Gyr, but  
its radius and strength are varying significantly. From $t\sim1$ Gyr on the bar grows almost linearly in time
to reach $R_{\rm bar}\sim 4\kpc$ at $t\sim9$ Gyr. Our definition for $R_{\rm bar}$ allows no exact comparison to the values
cited for the length of the MW long bar. \citet{wegg15} determine a length of 5 kpc and give 4 kpc as a lower limit, 
not in disagreement with our model.

\begin{figure*}
\centering
\vspace{-0.cm}
\includegraphics[width=18cm]{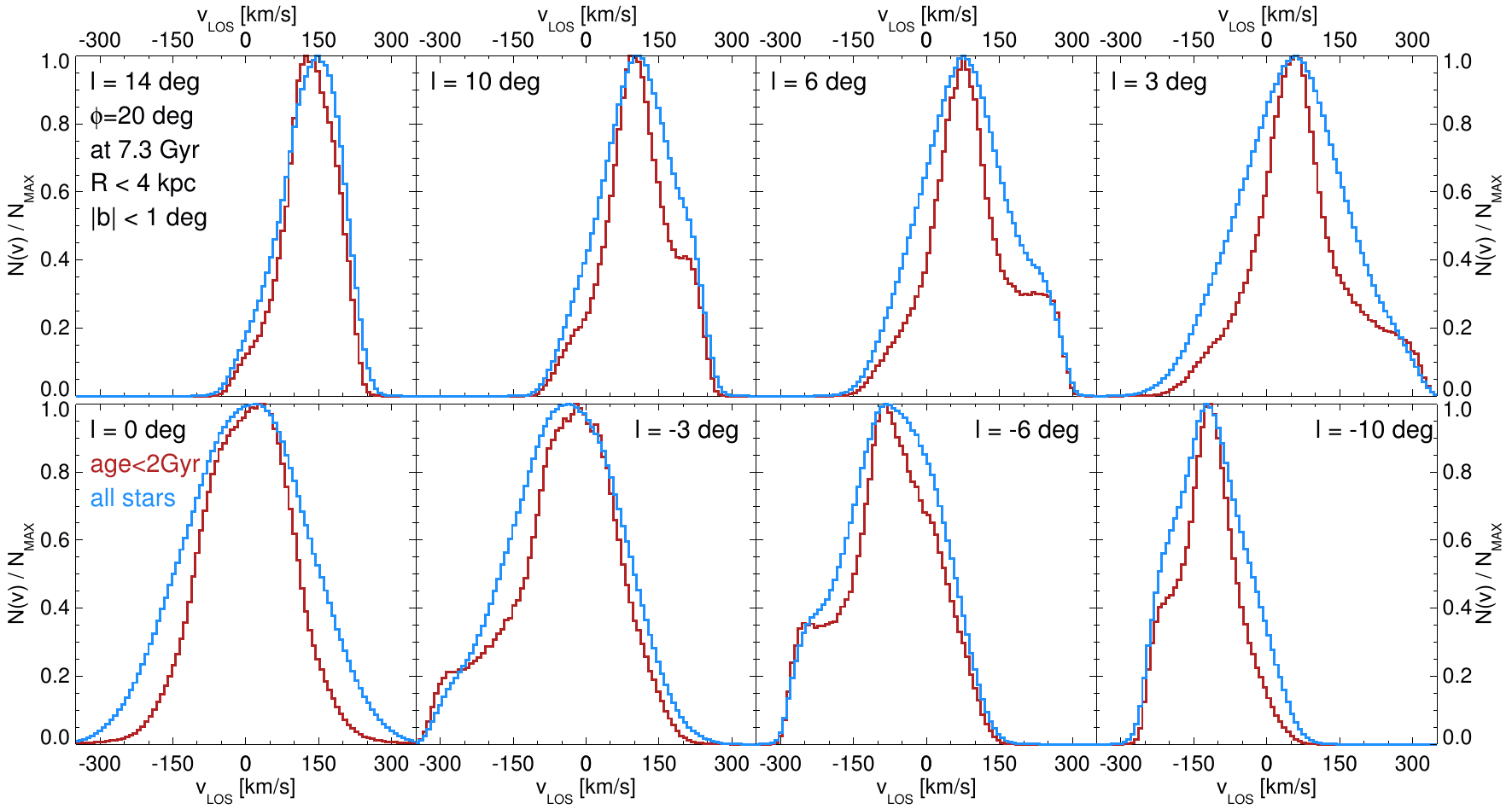}\\
\caption
{Line of sight velocity $\vlos$ distributions $N(\vlos)/N_{\rm max}$ of stars at $R<4\kpc$ as seen from $R=8.3\kpc$ at a bar angle of $\phi=20$ degrees
for eight different viewing longitudes $l=14-(-10)$ degrees. We consider stars at latitudes $\left|b\right|<1$ degrees and all panels cover 2 degrees width
in $l$. We show the distributions for all stars (blue) and stars younger than 2 Gyr (red). The $\vlos$ bins used here are all $8\kms$ wide.
}
\label{vlos-all}
\end{figure*}

In the middle panel of Figure \ref{bar-evo} we present the evolution of the bar rotational frequency $\Omega_{\rm bar}$ determined
by following the angle of the bar major axis from snapshot to snapshot. The bar initially rotates
with a frequency of $60-80 \kms \kpc^{-1}$, but as the loss of angular momentum to the halo is not balanced by the transfer of angular 
momentum  from the inflowing gas component, $\Omega_{\rm bar}$ steadily declines to reach $\sim 25 \kms \kpc^{-1}$ at the end of the simulation.
For the MW bar this quantity is little constrained. While our value is lower than most estimates for the MW \citep{dehnen, antoja, sormani},
the value is in concordance with a recent estimate by \citet{portail} (see their Section 8.3 for a discussion of
deduced MW bar pattern speeds). Our model is thus at the lower end of values inferred for the MW.  

In Figure \ref{images} we present surface density projections of the whole galaxy model and the bar region at a simulation time $t=7.3$ Gyr.
The lower panels show face-on views of the stellar and the gaseous components. Apparently, the galaxy is dominated by the bar within $R=4\kpc$
and shows flocculent spiral structure at outer radii. The upper panels give close ups of the bar region. The face-on view shows two distinct 
radial regions in terms of minor axis extent. At $R<2\kpc$, the bar shows a mildly elongated structure, at $R=2-4\kpc$ a thinner, long bar 
structure is visible. The edge-on view reveals, that the region $R<2\kpc$ is dominated by an X-shaped structure with the tips of this structure
at $(x,z)\sim(2,1.3)\kpc$, very similar to the structure of the MW bulge/bar region inferred by \citet{wegg13}. This view also shows that the 
outer bar is significantly thinner. Around the tips of the bar we measure a vertical exponential scale-height of 180 pc, similar to the long bar
scale-height inferred by \citet{wegg15}, but without a super-thin 45 pc component as found by these authors. Such a component is likely
too thin to be resolved in our simulation, considering our softening length of 30 pc.

At a solar neighbourhood like radius $R=8\kpc$, we find that our model at 7.3 Gyr has a stellar exponential scale-length of 3.5 kpc, which grows mildly towards 10 Gyr.
The disc is thus likely more extended than that of the MW, for which estimates usually range at 2-3 kpc. 
The model has a total baryonic mass of $4.2\times10^{10}\msun$ at 7.3 Gyr, it has grown from a mass of $0.5\times10^{10}\msun$ at $t=0$.

In the right panel of Figure \ref{bar-evo}
we plot the circular velocity curve $v_{\rm circ}(R)$ in the plane of the model galaxy at 7.3 and 9.3 Gyr. 
The contributions from dark matter and baryons are shown in red and blue. We measure $v_{\rm circ}=\sqrt{{a_{R}} R}$ 
by calculating the component of the gravitational acceleration pointing towards the centre ${a_{R}}$ on an $(R,\phi)$ grid at $z=0$ and at each $R$ average 
over the equally distributed values of $\phi$. We note that the circular velocity curve in the centre of the MW is not well constrained and that it
is uncertain how the observations of gas velocities in the MW bar are connected to our definition of $v_{\rm circ}(R)$.
There is only mild evolution from 7.3 to 9.3 Gyr visible in $v_{\rm circ}(R)$.

The circular velocity curves show features at the bar radius, deriving from the baryonic matter distribution. The reason for these features is that
the bar redistributes matter towards the centre, whereas by construction of the model new baryons are added outside the bar region, which gives rise to
the baryonic dip at $R\sim 3\kpc$. In our model, the baryonic contribution to the circular velocity is greater than the dark matter equivalent only within 
$R\sim 2.5\kpc$. It is unclear how this relates to the MW, but our model agrees with two important constraints:
A) At $R=8\kpc$ and 7.3 Gyr we recover $v_{\rm circ}=236 \kms$, in good agreement with what \citet{ralph} finds from local kinematics. 
B) The total dark matter mass within a sphere of $r=8\kpc$ is $6.3\times10^{10}\msun$ in good agreement with the dark matter mass within a sphere
of solar galactic radius that \citet{piffl} have deduced.

At the evolution stages relevant for this paper, the model has a slow bar with $R_{\rm bar}\sim 0.5 R_{\rm CR}$, where $R_{\rm CR}$ is the bar corotation radius.
This is likely connected to the fact that our galaxy model is {\it sub-maximum}, i.e. the dark matter contribution to the circular velocity curve is significant in the 
bar/disc region, as fast bars with $R_{\rm bar}= 0.7-1.0 R_{\rm CR}$ are thought to require {\it maximum} discs with baryon-dominated circular velocity curves \citep{debattista}.

\begin{figure*}
\centering
\vspace{-0.cm}
\includegraphics[width=18cm]{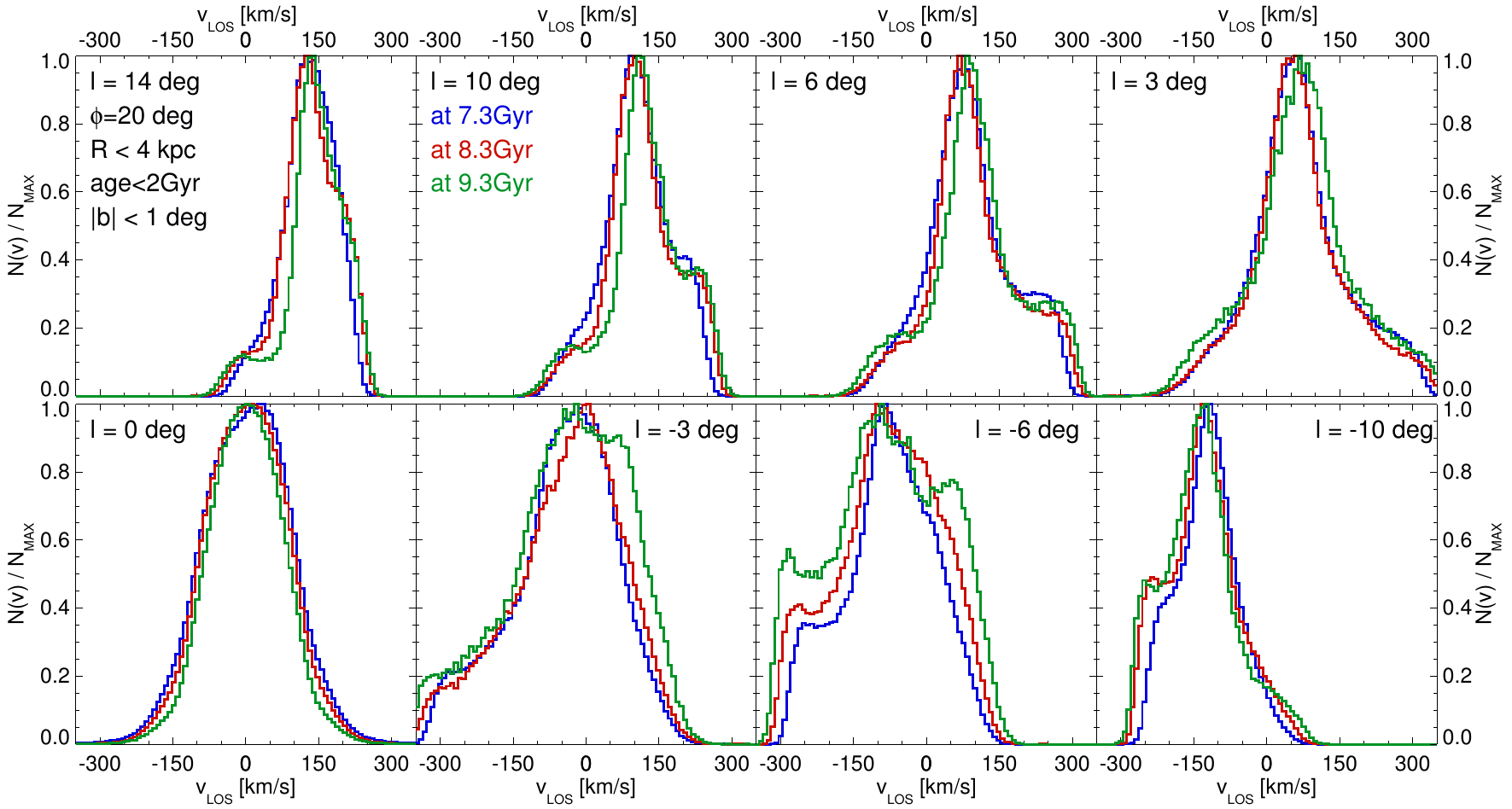}\\
\caption
{Line of sight velocity $\vlos$ distributions $N(\vlos)/N_{\rm max}$ of young stars with ages $<2$ Gyr at $R<4\kpc$ as seen from $R=8.3\kpc$ at a bar angle of $\phi=20$ degrees
for eight different viewing longitudes $l=14-(-10)$ degrees. We consider stars at latitudes $\left|b\right|<1$ degrees and all panels cover 2 degrees width
in $l$. We show the distributions for stars at three different times: 7.3 (blue), 8.3 (red) and 9.3 Gyr (green). The $\vlos$ bins used here are all $8\kms$ wide.
}
\label{vlos-young}
\end{figure*}

\section{$\vlos$ distributions}

In the previous section we have shown that the bar emerging in our galaxy model has comparable structural parameters to the MW bar.
We now extract line of sight velocity $\vlos$ distributions from our model and qualitatively compare them to the results
of \citet{nidever}. In this Section, we limit ourselves to the general understanding of the observed phenomena. We will  attempt a
quantitative comparison to observed data in the following Section.

\subsection{$\vlos$ distributions for the bar region}

As we expect stars moving along the major axis of the bar to be the source of the stars that comprise the high $\vlos$ peak, we focus
here on the central region of the model, defined by galactocentric  $R<4\kpc$. The full model
including the selection function will be discussed in the next Section.

To create $\vlos$ distributions we place ourselves at a solar radius of $R=8.3\kpc$ \citep{paul} and orient the bar to an angle of $\phi$
between its major axis and the Sun-Galactic centre sightline. Values found for $\phi$ in the literature range between 15 and 40 degrees.
For most of this paper we choose $\phi=20$ degrees and show in the next Section that our conclusions are unchanged for the full range of cited values.
We usually define sightlines which are 2 degrees wide in $l$ and $b$ and centred on the given values. To curb Poisson noise we use the
$\pi$ ($m=2$) symmetry of the model and use two sightlines per snapshots, 180 degrees apart in azimuth. For each plot we stack 50 snapshots, separated in time
 by 1 Myr and rotate the bar to achieve the same value of $\phi$ for each snapshot.

In Figure \ref{vlos-all}, we show $\vlos$ histograms at latitudes $\left|b\right|<1$ degrees at a simulation time of 7.3 Gyr for
eight different longitudes $l$ between $14$ and $-10$ degrees. The figure shows histograms for all bar stars with $R<4\kpc$
in blue and for the subset of young star particles with ages $<2$ Gyr in red. The $\vlos$ distributions for all stars (blue lines)
do not reveal significant features at high velocities. This is consistent with recent examinations of simulations in \citet{nidever}, 
\citet{li} and \citet{molloy}, who also detected no features when all stars are taken into account.

\begin{figure*}
\centering
\vspace{-0.cm}
\includegraphics[width=14.46cm]{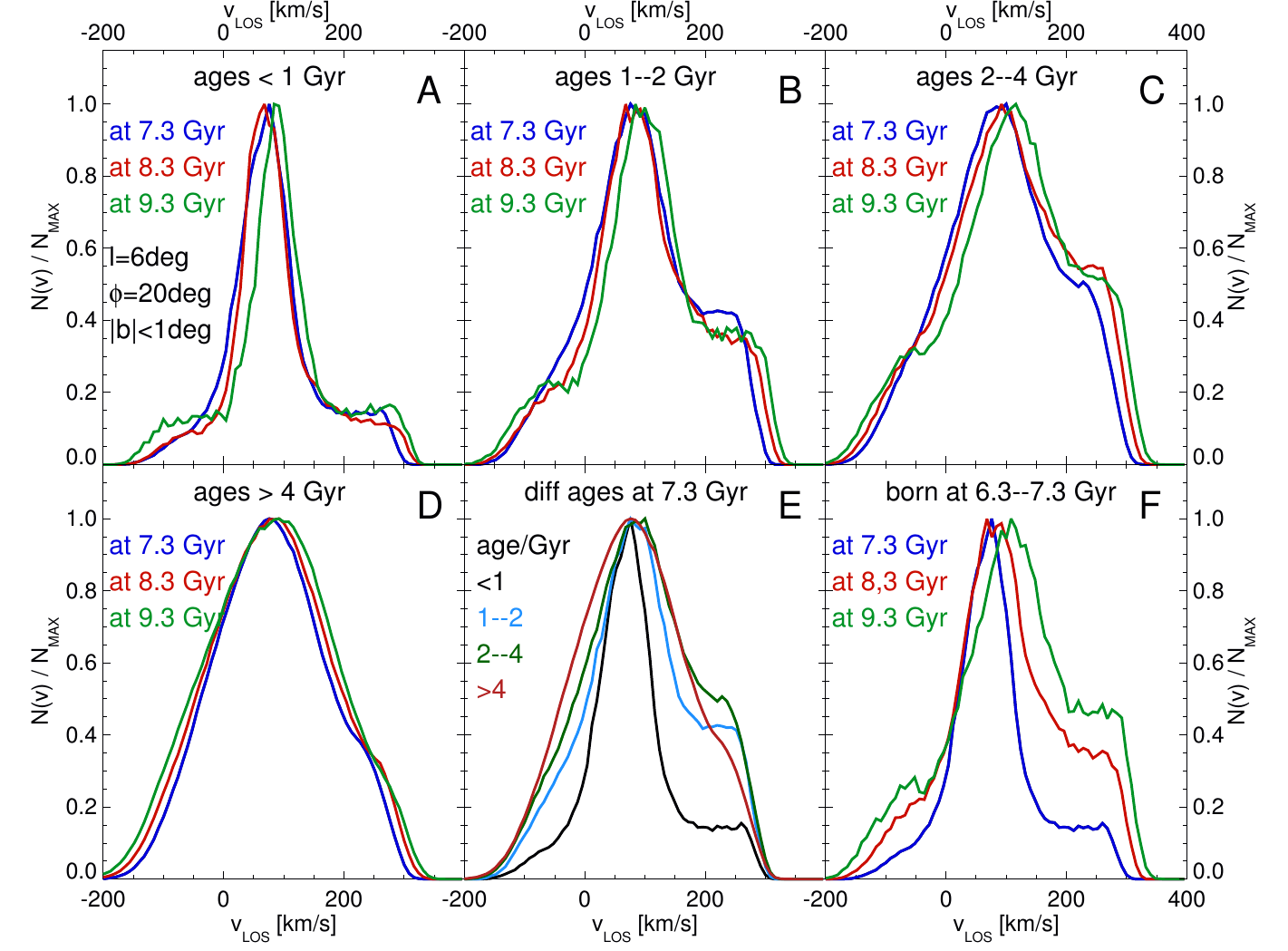}\\
\caption
{Line of sight velocity $\vlos$ distributions $N(\vlos)/N_{\rm max}$ of stars at $R<4\kpc$ as seen from $R=8.3\kpc$ at a bar angle of $\phi=20$ degrees
at longitudes $l=6\pm1$ degrees and latitudes $\left|b\right|<1$ degrees. $\vlos$ bins used here are all $8\kms$ wide.
{\it Panels A-D}: Populations of stars at four fixed age bins, $<1$ Gyr, 1--2 Gyr, 2--4 Gyr and $>4$ Gyr at three different
simulation times: 7.3 (blue), 8.3 (red) and 9.3 Gyr (green).
{\it Panel E}: Populations of stars at the four age bins $<1$ Gyr (black), 1--2 Gyr (blue), 2--4 Gyr (green) and $>4$ Gyr (red) at simulation time 7.3 Gyr.
{\it Panel F}: The population of stars born between 6.3 and 7.3 Gyr in the simulation as viewed at 7.3 (blue), 8.3 (red) and 9.3 Gyr (green).
}
\label{vlos-age}
\end{figure*}

\citet{nidever} reported peaks of high velocity stars
with $\vlos\sim 200-250\kms$ of varying extent for longitudes $l$ between $14$ and $4$ degrees and \citet{babusiaux} reported such peaks
for $l=10$ and $6$ degrees and a peak at $\vlos\sim -225\kms$ at $l=-6$ degrees.  
Closer inspection of the blue histograms at $l=\pm 6$ and $\pm 10$ degrees reveals an excess of stars at the expected $\vlos$ values.
Moreover, the position of the main peak in the $\vlos$ distribution decreases with decreasing $l$ as observed by \citet{nidever}. These main peaks of the
distributions are, however, wider in the model than in the observation. 

The fact that the APOGEE selection function (Figure \ref{selfu}) favours young stars motivates to examine only stars with ages $<2$ Gyr
shown as red curves. Note that stars formed from low angular momentum central gas do not play a role for this analysis and
are excluded here. The main $\vlos$ peaks are much narrower for young stars than for all stars, indicating that these populations are cooler.
The red histograms clearly show distinct components at $\vlos=200-300\kms$ for $l=3,6$ and $10$ degrees
and at $\vlos=-300-(-200)\kms$ for $l=-3,-6$ and $-10$ degrees, in agreement with observations.  Interestingly, we do not recover a separate
component at $l=14$ degrees, where \citet{nidever} also claimed to find a distinct high $\vlos$ peak, a finding which we will return to in the next Section.

\subsection{Evolution with time and age}

As the model young populations show the observed signature, it is interesting to ask whether this is specific to a unique time or stage of bar evolution
or a generic feature of young stars. In Figure \ref{vlos-young} we examine the evolution of young stellar populations with time plotting the populations
which at respective times are less than 2 Gyr old. We consider simulation times  7.3, 8.3 and 9.3 Gyr. We notice several points:
i) There is very little change in the general shape of the histograms and in the position and width of the main $\vlos$ peaks at all longitudes $l$. 
ii) The extreme values of $\vlos$ increase with time by around $20\kms$ for all sightlines which do not cross
the galactic centre. iii) The high $\left|\vlos\right|$ peaks identified in Figure \ref{vlos-all} become slightly more distinct in time and 
also move to higher velocities. iv) For 9.3 Gyr additional features/peaks become apparent at the opposite side of the main 
peaks, e.g. at $-150<\vlos/\kms<-50$ at $l=6$ degrees or at $0<\vlos/\kms<100$ at $l=-6$ degrees.

The high $\vlos$ features at positive longitudes and low $\vlos$ features at negative longitudes are thus stable over time and not connected to any temporary phenomenon.
Due to the lack of opposite peaks in the APOGEE data we will focus our analysis on the simulation time 7.3 Gyr.

So far we have shown that young stars show high $\left|\vlos\right|$ peaks, whereas the whole population of bar stars does not. It is thus interesting  to study in more 
detail how the $\vlos$ distribution evolves with age. We attempt this in Figure \ref{vlos-age}, where we concentrate on one sightline, $l=6\pm1$ degrees, for which both
our simulation and APOGEE data display a peak at high $\vlos$.

Panel E shows how the $\vlos$ distribution varies when stars are sorted in four different age bins. Whereas the high $\vlos$ population
is very distinct for stars with ages below 1 Gyr, this population becomes less distinct at higher ages. For stars with ages between 1 and 2 Gyr, it is still clearly visible,
at 2-4 Gyr only a small feature remains and for older stars even this weak signature vanishes. The vanishing of the distinct 
high $\vlos$ population goes hand in hand with broadening of the main $\vlos$ peak indicating a heating of the population.

\begin{figure*}
\centering
\vspace{-0.cm}
\includegraphics[width=14.46cm]{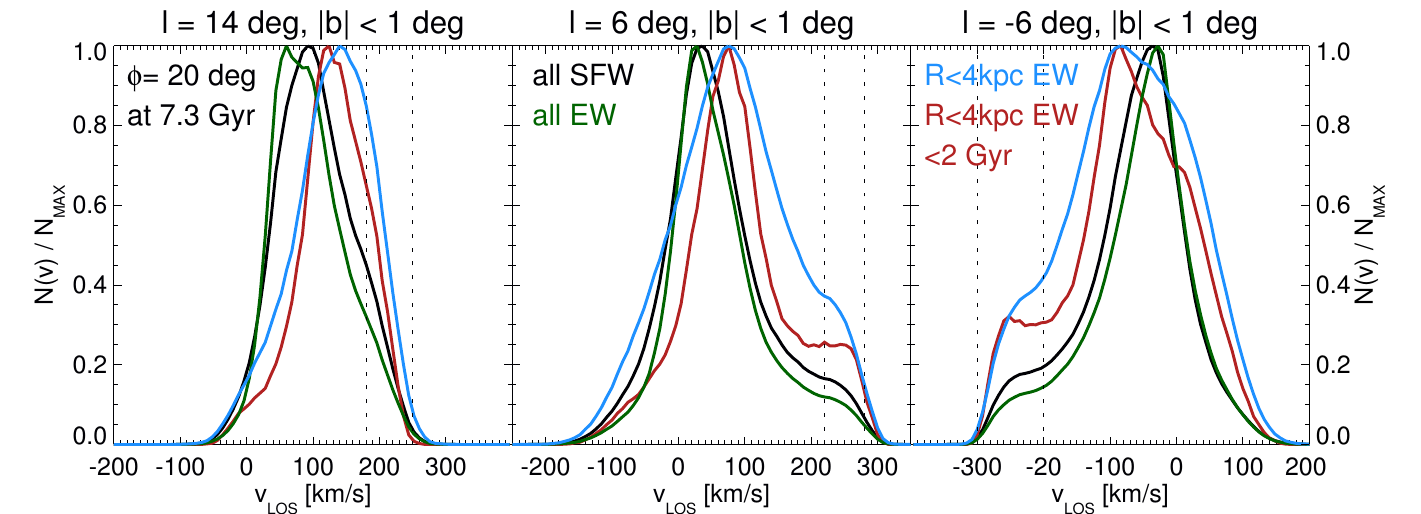}\\
\caption
{Line of sight velocity $\vlos$ distributions $N(\vlos)/N_{\rm max}$ of stars at longitudes $l=14\pm1$, $l=6\pm1$ and $l=-6\pm1$ degrees and latitudes $\left|b\right|<1$ degrees.
 We assume a bar angle of $\phi=20$ degrees, a solar radius of $R=8.3\kpc$ and use $\vlos$ bins which are all $8\kms$ wide. We compare four different types of histograms:
{\it blue} considering all stars at $R<4\kpc$, equally weighted (EW), 
{\it red} considering young stars with ages $<2$ Gyr at $R<4\kpc$, equally weighted, 
{\it green} considering all stars along the sightline, equally weighted and
{\it black} considering all stars along the sightline, weighted according to the age- and distance-dependent selection function ${\rm SF}(s,\tau)$ from Figure \ref{selfu} (SFW).
The vertical dashed lines indicate the intervals for peak stars considered for Figure \ref{sightline}.
}
\label{compare}
\end{figure*}

To see how this evolves with simulation time, we plot in Panels A-D stars in these age bins at three different simulation times, 7.3, 8.3 and 9.3 Gyr.
We notice that there is much less evolution with simulation time than there is with age. The level of distinction of the high $\vlos$ signature remains similar at all considered
simulation times for each age bin.

To strengthen these findings we plot in Panel F of Figure \ref{vlos-age} the population of stars born between 6.3 and 7.3 Gyr at three different times
representing mean population ages 0.5, 1.5 and 2.5 Gyr. This underlines that the evolution of the $\vlos$ distribution with age shows the broadening of the main $\vlos$ peak
due to heating and the gradual loss of the high $\vlos$ peak.

We again notice that the young populations of stars at 9.3 Gyr show additional features at the opposite side of the main peak.
Closer inspection reveals hints of this feature also at earlier times. Additionally, the main peaks move to higher $\vlos$. 
We will provide an explanation for these changes in Section 7. We note, however, that these additional peaks are only visible if we select young bar stars.
When applying the selection function on all stars along the line of sight, as described in the next Section , these features, unlike the conventional 
high $\vlos$ peaks, are too small to be detected.

\section{Comparison to APOGEE data}

In the previous Section, we have shown that populations of young bar stars are expected to display high $\left|\vlos\right|$ features as observed by \citet{nidever} and 
\citet{babusiaux}. In reality, not only stars with $R<4\kpc$ will contribute to the APOGEE $\vlos$ distributions, but also disc stars in front of or behind the bar.
In the following, we directly compare the $N$-body simulation to the APOGEE data, applying the selection function described in Section 2.

\begin{figure*}
\centering
\vspace{-0.cm}
\includegraphics[width=16cm]{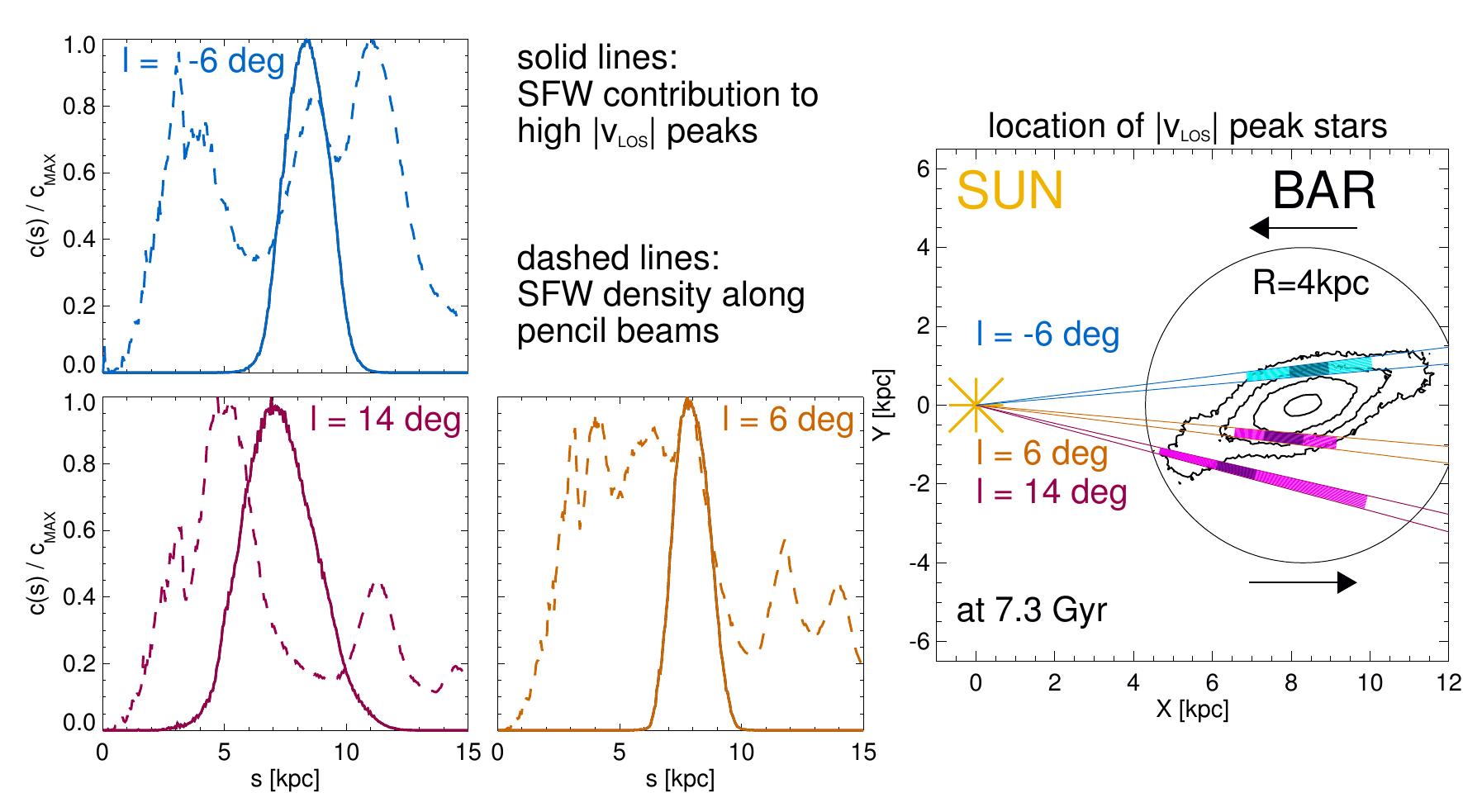}\\
\caption{{\it Left panels}: The selection function weighted distribution of all stars in a sightline (dashed lines) and of the stars contributing to the high $\avlos$ peaks (solid lines)
                       as a function of their distance from the Sun $s$, $c(s)/c_{\rm max}$. We show longitudes $l=14\pm1$ (red), $l=6\pm1$ (orange) and $l=-6\pm1$ (blue) degrees. 
                       High $\avlos$ windows are defined as $180<\vlos/\kms<250$ for $l=14\pm1$, $220<\vlos/\kms<280$ for $l=6\pm1$ and $-300<\vlos/\kms<-200$ for $l=-6\pm1$, 
                       as indicated in Figure \ref{compare}.
{\it Right panel}: The location of the high $\avlos$ peak stars in the $x-y$ plane. We show the general geometry of single sightlines from the Sun (yellow star) into the bar 
                   (stellar mass density contours from Figure \ref{images} in black). Arrows indicate galactic and bar rotation. The black circle marks $R=4\kpc$.
                   The magenta areas show the distance areas which contain 95 per cent of the SFW stars contributing to the high $\left|\vlos\right|$ peak windows
                   for stars moving away from the sun, cyan areas are for high $\avlos$ stars moving towards the sun at negative longitudes. Darker shaded areas 
                   show the areas of $\pm0.5\kpc$ centred on the peaks of $c(s)$. Note the position of the fast moving stars with $180<\vlos/\kms<250$ at $l = 14$,
                   which indicates the increasing contribution of stars outside the bar.
}
\label{sightline}
\end{figure*}

\subsection{The effect of the selection function}

We start with Figure \ref{compare}, where we compare the $\vlos$ histograms for stars with galactocentric radii $R<4\kpc$ (see previous Section) to $\vlos$ histograms obtained
for the full sightlines. We do so for three longitudes, $l=\pm6$ and 14 degrees. The blue curves represent all stars within 4 kpc and the red curves show the subset of young stars, both are taken from
Figure \ref{vlos-young}. The velocity distributions for all stars along the sightline (equally weighted, green) are strikingly different. Their main $\vlos$ peaks
are significantly narrower and shifted by more than $50\kms$ towards zero.
This is due to the dominant contribution by fore- and background disc stars, which are kinematically cold and 
are observed with the local azimuthal direction nearly perpendicular to the line of sight.

We noticed above that no distinct high $\left|\vlos\right|$ feature is found for all stars at $R<4\kpc$. Interestingly this is not true for all stars along the
line of sight. In this population, the feature is less distinct than the one for young bar stars, but lies at the same $\vlos$ range.
The histograms obtained when weighting all stars along the line of sight according to the selection function presented in Section 2 are shown as black lines. 
Overall, they resemble more the ones for all stars along the line of sight than to the ones for stars at $R<4\kpc$. When present, the high $\left|\vlos\right|$ feature
is strengthened by the selection function and the main peak is shifted to slightly larger $\left|\vlos\right|$.

From these comparisons we conclude that the $\vlos$ distribution is not only shaped by stars in the bar region, but by all stars along the line of sight.
This is illustrated by the selection function weighted (SFW) distribution of stars along the line-of-sight $c(s)$ as a function of distance from the sun $s$ in the left panels
of Figure \ref{sightline} (dashed lines). Clearly, the SFW distribution functions do not simply reflect the density distributions along the pencil beams,
but are shaped by the age and distance dependent selection function modulated by the beam area increasing as $s^2$. Of the three shown sightlines, only the one which crosses
the bar most centrally ($l=6$ degrees) has its absolute maximum coincident with the bar distance. The loci of young stars, the highest densities of which are by construction at a
galactocentric radius $R\sim4\kpc$ and thus at distances $s\sim 4$ and $12\kpc$ are visible as local maxima. Moreover, spiral structure creates minor peaks.

The high $\vlos$ feature appears connected to young bar stars. To verify this assumption, we plot in the left panels of Figure \ref{sightline} the contribution to
the high $\avlos$ features of SFW stars for the three sightlines of Figure \ref{compare} as a function of distance from the Sun $s$ (solid lines). The intervals we use as definitions for these features
are indicated by dotted vertical lines in Figure \ref{compare}. For $l=\pm6$ degrees, we were guided by the distinctly visible feature, for $l=14$ degrees, 
where we find no feature, we use the estimate by \citet{nidever}.

\begin{figure*}
\centering
\vspace{-0.5cm}
\includegraphics[width=16.5cm]{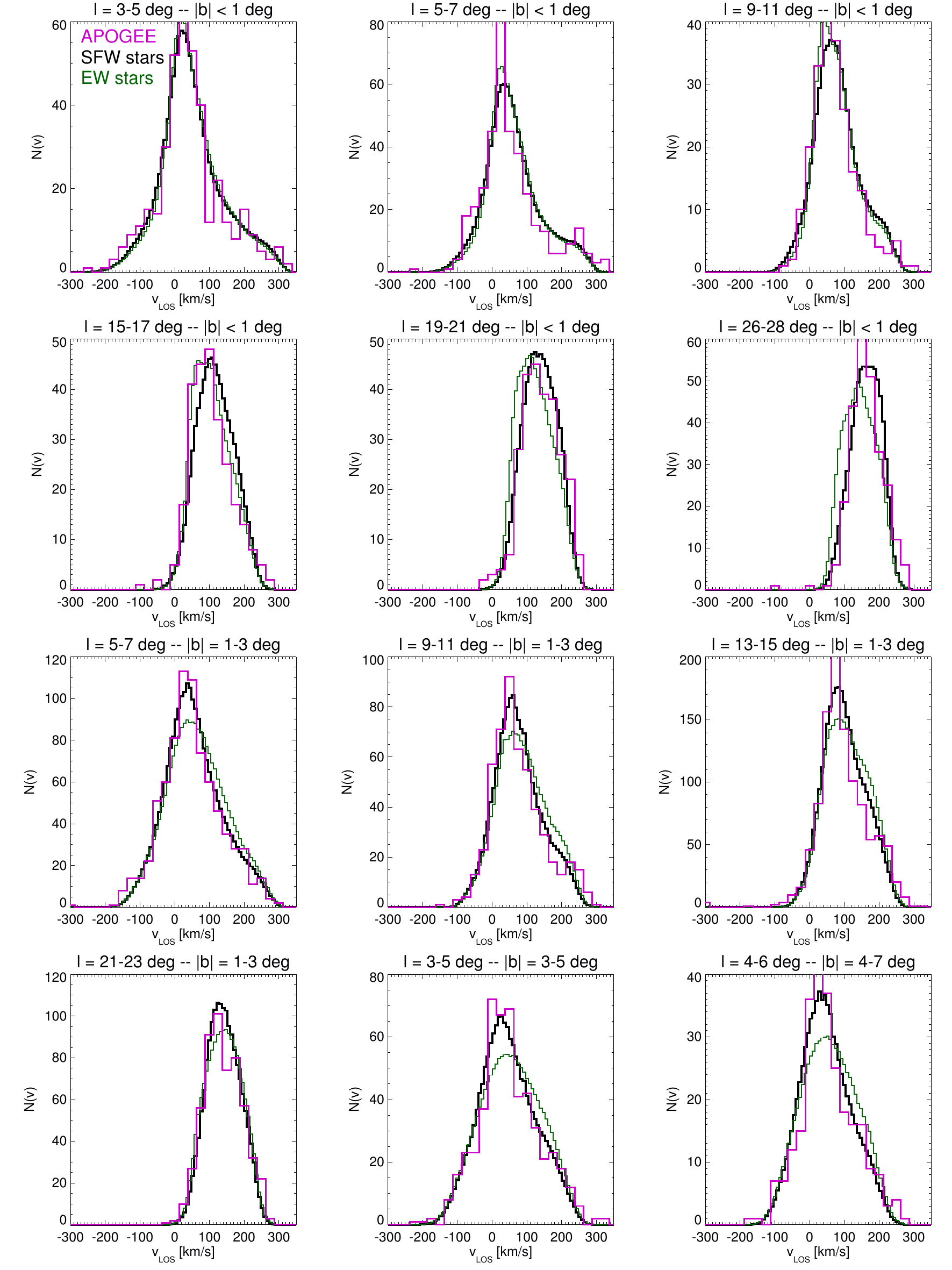}\\
\caption
{A comparison with APOGEE data. For 12 sightlines for which data from APOGEE are available we compare line of sight velocity $\vlos$ distributions $N(\vlos)$
to the data discussed in Section two. The APOGEE data shows the actual number of stars per bin, the simulation data are normalised to yield the same total number.
The APOGEE $\vlos$ bins are $25\kms$ wide, whereas the simulation bins are $8\kms$ wide. For comparison to APOGEE data (magenta), 
we show simulation histograms for equally weighted (EW, green) and selection function weighted (SFW, black) stars along the sightline. We assume a bar angle of $\phi=20$
degrees and a solar radius of $R=8.3\kpc$.
}
\label{apogee}
\end{figure*}

At each sightline in Figure \ref{sightline} the regions of contributions to the high $\avlos$ stars are much narrower than the SFW distribution of all stars in the pencil beam.
Moreover, the maxima of contribution to the high $\avlos$ stars are (apart from $l=6$ degrees) offset from the maxima of the SFW distribution of all stars.
The distance distribution of contributing stars $c(s)$ is narrowest for $l=6$ degrees and widest for $l=14$ degrees. The main peaks are at $s=7-8.5\kpc$ on
the rear side of the bar (front side for $l=-6$ degrees).
To visualise where these contributions arise in relation to the bar, we use the right panel of Figure \ref{sightline}, where we show the position of the Sun in yellow,
the considered sightlines in blue, red and orange and the surface density contours of the stellar mass in the bar region at 7.3 Gyr in black (see Figure \ref{images}).
The 95 per cent contribution regions of high $\avlos$ stars for each sightline are shown in magenta for stars moving away from the sun and in cyan for stars moving towards the sun at negative longitudes.
We highlight the regions of $\pm0.5\kpc$ around the peak distances in darker shades. For $l=\pm6$ degrees, these regions are fully within the bar region
and strictly on one side of the bar (rear side for $l > 0$, front for $l < 0$), as the fast stars travelling alongside 
the bar have to lose angular momentum on the way in and gain it on the way out. The contribution region is narrower for $l=6$ degrees as
 this sightline crosses the bar at a greater angle than the one for $l=-6$ degrees.

This indicates that the high $\left|\vlos\right|$ features found by \citet{nidever} correspond to stars which are associated with the
bar region and not with disc orbits. This confirms the ideas of \citet{nidever} and \citet{molloy} and rejects the notion
of \citet{li} that the observed features are not related to the bar.

At $l=14$ degrees there is a tentative feature in observed data, but not in our simulations. If we select the highest $\vlos$ stars in the simulation we find that the main contributions 
come from the edge of the bar, but extend far into the disc region behind the bar. At these longitudes, the inner parts of the disc align with the line of sight
and start to contribute to the high $\vlos$ window. Should the high $\vlos$ feature at $l=14$ degrees prove to be real, a clear attribution may only be possible with detailed
analytical models.

\begin{figure*}
\centering
\vspace{-0.cm}
\includegraphics[width=18cm]{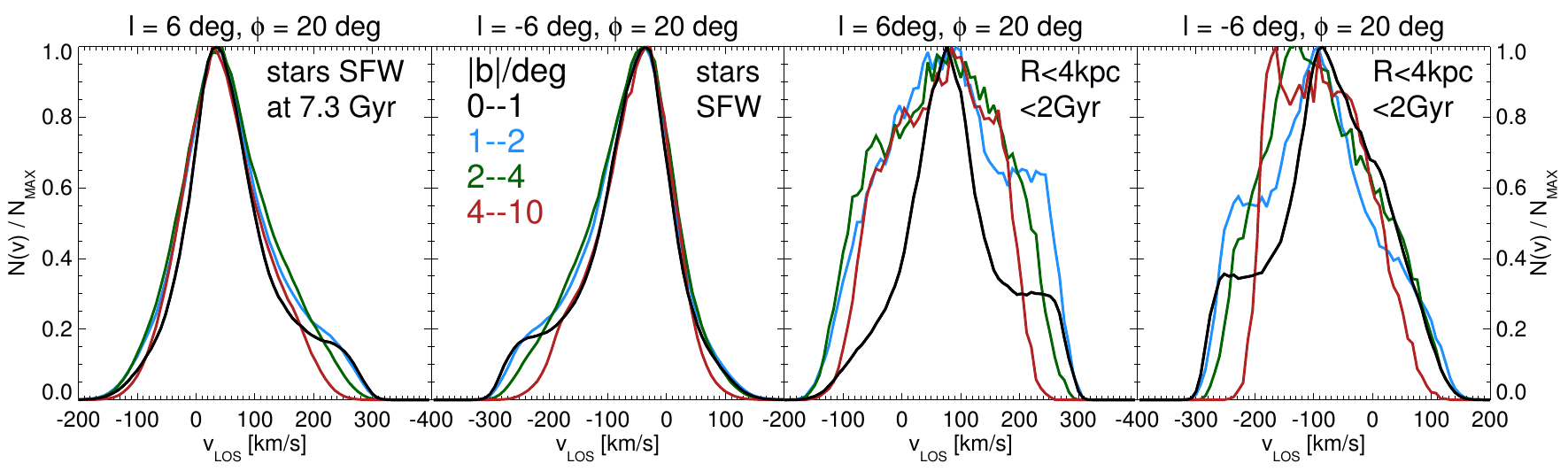}\\
\caption
{Line of sight velocity $\vlos$ distributions $N(\vlos)/N_{\rm max}$ as seen from $R=8.3\kpc$ at a bar angle of $\phi=20$ degrees at a simulation time of 7.3 Gyr.
 $\vlos$ bins used here are all $8\kms$ wide. We consider sightlines centred on $l=\pm6$ degrees, which are both 2 degrees wide in $l$.
 In each panel we consider four latitude bins: $\left|b\right|<1$ degrees (black), 1--2 degrees (blue), 2--4 degrees and $>4$ degrees (red).
{\it Left panels}: Here we consider histograms of all stars along the sightline, weighted by the selection function depicted in Figure \ref{selfu}.
{\it Right panels}: Here we consider equally weighted, young stars (ages $<2$ Gyr) at $R<4\kpc$.
}
\label{diff-b}
\end{figure*}

\subsection{Detailed comparison with APOGEE}

We now directly compare our model with APOGEE data. Remember that we have done nothing at all to fit this specific model galaxy to the MW. 
There may be significant differences in length, mass, age and detailed history of the bar etc.
Hence the goal is to qualitatively identify features and not to do a full quantitative analysis.

In Figure \ref{apogee} we compare $\vlos$ histograms for all stars along the sightlines, equally weighted (EW, green) or selection function weighted
(SFW, black) to the APOGEE data described in Section 2 for various sightlines. 
We display in-plane data at $\left|b\right|<1$ degrees in the upper two rows, and sightlines at higher $\left|b\right|$ in the bottom two rows.
Despite its limitations, the simulation fits the data surprisingly well. The SFW models
provide a better fit to APOGEE data than the EW models. Their high $\vlos$ features are more distinct and the EW histograms are 
offset from the data at large $l$ as well as at large $b$.
 
The sightlines with $l\gtrsim 14$ degrees are increasingly dominated by disc stars. There, the parts aligned with the line of sight 
would bury any distinct velocity feature from the bar. In concordance with this expectation, 
there is no discernible feature in the corresponding in-plane APOGEE data, and also not in the model. In contrast, at $l=3-10$ degrees, the high velocity feature is prominent both in the MW data and our simulation. 
Taking into account the significant Poisson noise in the observations, there is little difference at $l\le6$ degrees.
At $l=10$ degrees, the feature is less distinct in the simulation. Interestingly, at $l=16$ degrees the SFW histogram provides
a worse fit than the EW histogram, as it misses the position of the main peak, which at other longitudes is well reproduced. 

When we look slightly above and below the plane at $\left|b\right|=1-3$ degrees, asymmetries at high $\vlos$ are present in APOGEE data at $l=5-15$ degrees,
but again not at greater $l$. The tentative $l=14$ peak in the data was highlighted by \citet{nidever}, but has a very weak counterpart in the simulation.
The SFW model histogram here drops less steeply than the data on the high velocity side of the main peak, as is the case for $l=16$ degrees (mentioned above).
These are the strongest discrepancies between model and observations. There may be various explanations for this:

i) We know that the Galactic potential will not be perfectly matched. In particular for disc contributions, the location of the main
$\vlos$ peak is determined by the radial potential gradient (in idealisation the circular speed). If this differs, simulation and data will show a systematic
difference. ii) We know that the structure in particular in this region will show differences between simulation and observed data, e.g. caused by a difference
in bar length. A longer bar would mean that we see regions further away from the bar tips. We can achieve such an effect also by lowering the solar radius. Indeed,
if we do so, the asymmetry becomes more prominent. iii) Another explanation could be the dense ring of molecular clouds at $R = 4\kpc$ in the MW \citep{Dame87}.
This ring is associated with an intense radial peak in star formation, which is not matched by our simulations. The selection function of APOGEE is very sensitive to young stars. 
This feature will have no strong consequence when the sightline is perpendicular to circular orbits at $R \sim 4 \kpc$ (at small $l$) or outside the region ($l\gtrsim30$),
but sightlines at $l\sim 16$ degrees will be strongly affected. As they pass through $R = 4 \kpc$ at an intermediate angle to the local azimuthal direction,
the enhancement of these stars will drive up stellar densities at intermediate $\vlos$ and shift the main peak.
Overall, the complicated contribution function  of stars at these sightlines prevents any strong conclusions to be drawn from this discrepancy.

Moving even further away from the plane, we consider latitudes $\left|b\right|\ge3$ degrees in the lower row of Figure \ref{apogee}. There is no discernible feature, 
either in observations or simulation. The slight asymmetry is linked to the rotation of the bar. So our model provides reasonable fits to the data in and out of the plane.

\begin{figure*}
\centering
\vspace{-0.cm}
\includegraphics[width=14.46cm]{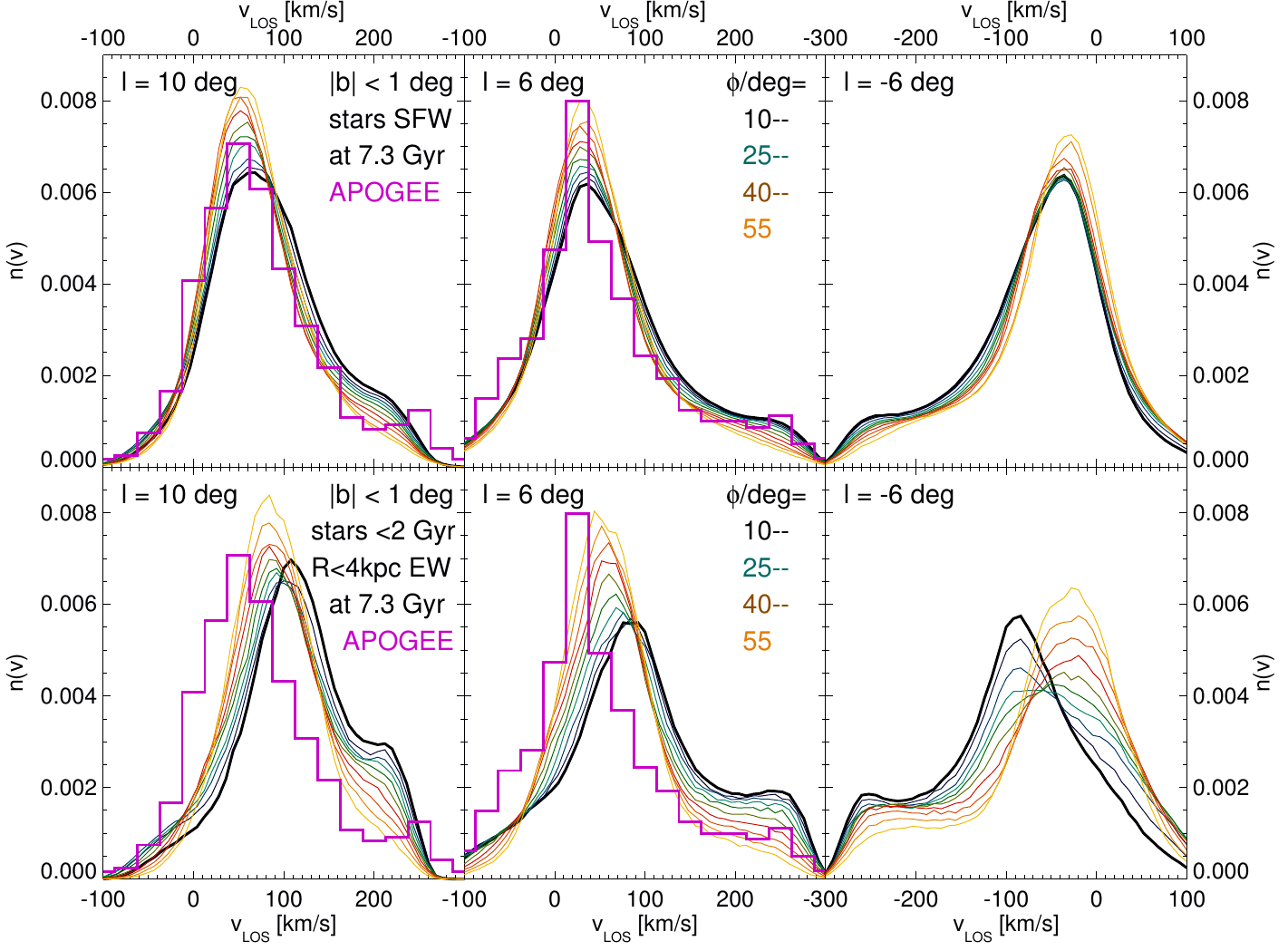}\\
\caption
{Normalised line of sight velocity $\vlos$ distributions $n(\vlos)$ of stars at 7.3 Gyr simulation time and latitudes $\left|b\right|<1$ degrees assuming a solar radius $R_{\rm sun}=8.3\kpc$.
  If available APOGEE data are overplotted in magenta. In each panel we consider bar angles $\phi$ between 10 and 55 degrees.
{\it Upper row}: All stars in the sightline, weighted by the selection function presented in Figure \ref{selfu}.
{\it Lower row}: All stars within $R<4\kpc$ and with ages $<2$ Gyr, equally weighted.
{\it Left column}: Longitude $l=10\pm1$ degrees. {\it Middle column}: Longitude $l=6\pm1$ degrees. {\it Right column}: Longitude $l=-6\pm1$ degrees. 
}
\label{phi}
\end{figure*}

\subsection{Varying the parameters}

We now examine the behaviour at different latitudes. Figure \ref{diff-b} shows simulation data for two sightlines $l=\pm6$ degrees and varying latitude. We do this both
for SFW histograms for all stars along the lines of sight (left panels) and for EW young central stars (right panels). 
The difference between these two types of histograms is even more striking at higher $\left|b\right|$ where for young central stars they are significantly broader.
This indicates that young bar stars play a small role in observations at higher latitude. Note also that the number of young bar stars decreases strongly with 
increasing latitude in the simulations, which explains the noisier nature of the corresponding histograms. We find that the exponential scaleheight $h_z$
for stars with ages $<2$ Gyr, which contribute to the high $\vlos$ features is $h_z\sim70-80\pc$, compared to $h_z\sim 180\pc$ for all stars at the ends of the bar.

All panels show that distinct high $\left|\vlos\right|$ features
are only present close to the plane. At $\left|b\right|=1-2$ degrees a mild feature is still present, while at greater $b$ there is nothing left.
This explains why \citet{zoccali} did not discover such features in their data, which are mostly at $\left|b\right|\ge2$ degrees.
Their only field at smaller latitude is at $l=0$, where we do not expect a peak anyway. This can be explained by the fact that the high $\vlos$ populations are dynamically cool
and thus also live at small vertical distances from the galactic plane. 

So far, we have assumed a bar angle of $\phi=20$ degrees. In the literature, estimates range from 15 to 40 degrees. \citet{molloy} recently claimed that the
high $\vlos$ features observed by \citet{nidever} could help to constrain $\phi$ and would favour low values $\phi\sim 15$ degrees. In Figure \ref{phi}, we thus attempt to
study the influence of $\phi$ on $\vlos$ distributions. We consider angles between 10 and 55 degrees for three sightlines
and both SFW stars (upper row) and young stars with $R<4\kpc$ (lower row).

\begin{figure*}
\centering
\vspace{-0.cm}
\includegraphics[width=14cm]{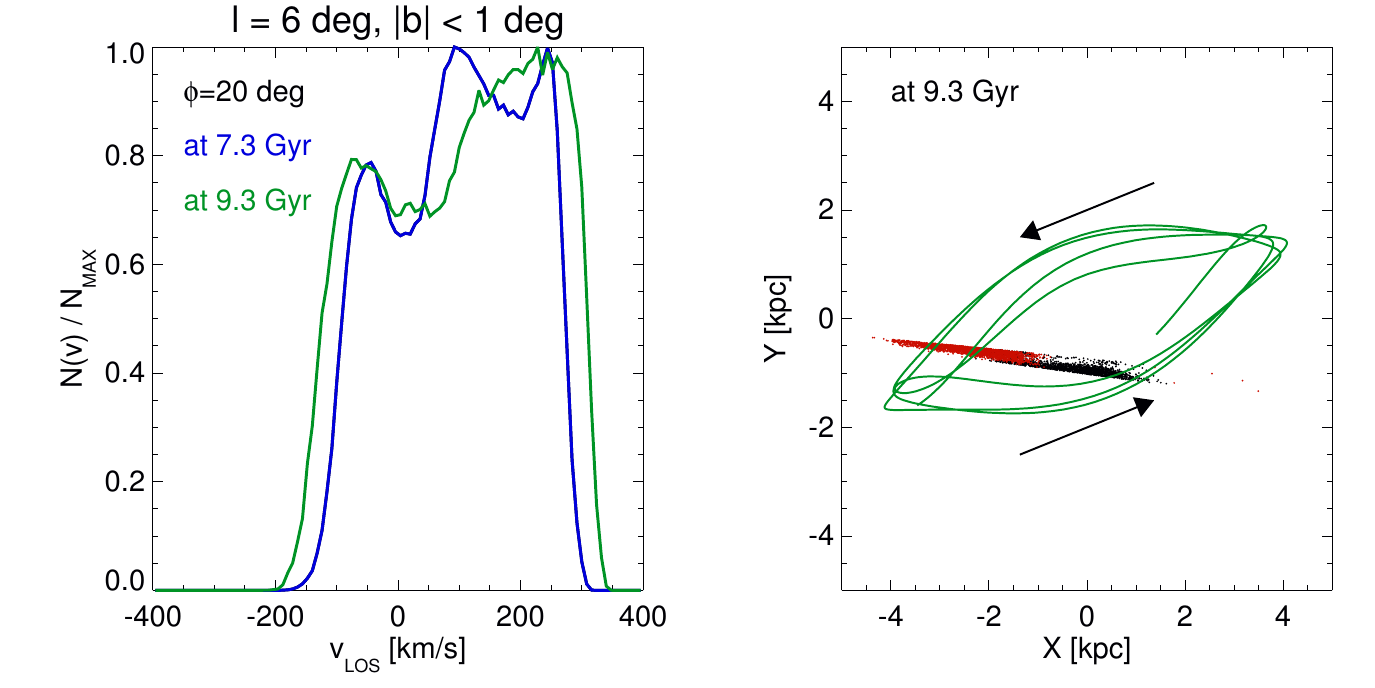}\\
\caption
{{\it Left panel}: Line of sight velocity $\vlos$ distributions $N(\vlos)/N_{\rm max}$ of young stars at $R<4\kpc$ as seen from $R=8.3\kpc$ with a bar angle $\phi=20$ degrees
                   at longitudes $l=6\pm1$ degrees and latitudes $\left|b\right|<1$ degrees. The stars were selected as members of the high velocity peaks, defined 
                   by $210<\vlos/\kms<280$ at 7.30-7.35 Gyr (blue) and by $230<\vlos/\kms<300$ at 9.30-9.35 Gyr (green). Then they were followed for 500 Myr with a snapshot separation 
                   of 1 Myr and each time they were in the sightline $l=6\pm1$, their $\vlos$ was recorded to produce the shown $\vlos$ distributions.
{\it Right panel}: The positions of stars at 9.30-9.35 Gyr seen at longitudes $l=6\pm1$ degrees and latitudes $\left|b\right|<1$ degrees and either in the high $\vlos$
                   peak ($230<\vlos/\kms<300$, black dots) or the low $\vlos$ peak ($-150<\vlos/\kms<-50$, red dots). Each star is recorded only once, 
                   on its first detection in the corresponding velocity window. The sun is at $x=-8.3\kpc$ and $y=0$ here.
                   Overplotted in green is an orbit over 500 Myr, which during this time alternately contributes to both peaks. The arrows indicate the sense of motion.
}
\label{vlos-peak}
\end{figure*}

We see that both for positive and negative $l$, the high $\left|\vlos\right|$ features are less distinct for large $\phi$. 
This can be explained by the larger angle between the sightline and the bar. As the high velocity stars 
move along the bar, their projected $\vlos$ decreases. At larger $\phi$, stars on different orbits moving at greater angles to the bar can be
seen at higher $\vlos$. As they are less numerous, the feature becomes less prominent. This is especially clear for young stars at $R<4\kpc$.
If one tried to fit multiple Gaussians to the $\vlos$ distribution, as did \citet{nidever} and \citet{molloy}, one would find that the high $\vlos$ peak
becomes broader and centred at lower $\vlos$ as reported by \citet{molloy}.

In the upper panels of Figure \ref{phi}, APOGEE data are plotted over the SFW model. We should keep in mind the uncertainties in the observations
and the fact that the structure and age distribution of the model bar differ in an unknown extent from the MW bar. We thus conclude that the APOGEE data put only mild
constraints on the bar angle. They favour small angles, but only strongly disfavour angles $\phi\gtrsim40$ degrees.

The estimate by \citet{molloy} is also intrinsically overly confident, as they ascribe the high $\vlos$ feature to one orbit family only, whereas, as we will
show in the next Section, several families can contribute. Moreover, degeneracies with the potential, the pattern speed, the bar structure etc complicate
their determination of $\phi$.

\section{Tracking Orbits}

To understand better the origin of the high and low $\vlos$ peaks discussed above, we focus in this section on the orbits producing these features.

\subsection{High and low $\vlos$ features}

We created snapshots of the simulations at a frequency of 1 per 1 Myr from simulation time 7.3 Gyr onwards. We first consider the
following exercise: What would the $\vlos$ distribution look like if it were only made up of stars that at some point contribute to the high $\vlos$ peak.
We therefore select all stars with ages $<2$ Gyr and $R<4\kpc$ that at 7.30--7.35 Gyr contribute to the high $\vlos$ peak at $l=6\pm1$ degrees and $\left|b\right|<1$ degrees.
We then follow these particles for a further 500 Myr and construct a $\vlos$ distribution at the same $l$ and $b$ ranges only from these stars.
We do the same 2 Gyr later for a new set of stars selected in the same way. 
Figures \ref{vlos-age} and \ref{vlos-young} have already shown that the maximum (minimum) $\vlos$ observed increase (decrease) by about $20\kms$ within these 2 Gyr.
This is because of a growth of both the potential well and the bar and also because of the slowdown of the bar which allows
orbits to have lower angular momentum and reach positions closer to the centre and thus cover a larger potential difference.
Consequently, we increase the high $\vlos$ window from $210<\vlos/\kms<280$ to $230<\vlos/\kms<300$.

\begin{figure*}
\centering
\vspace{-0.cm}
\includegraphics[width=18cm]{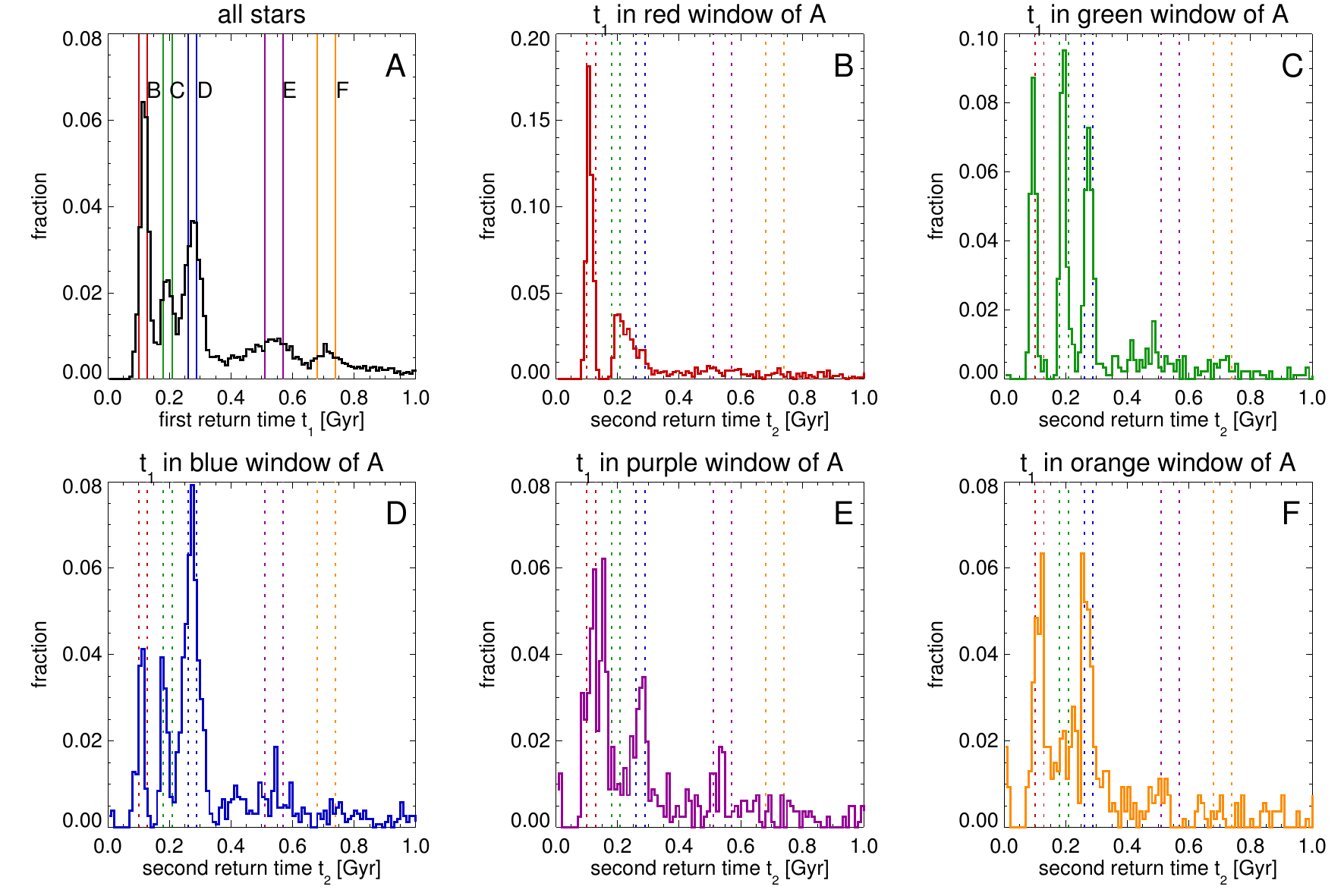}\\
\caption
    {For this Figure, we track stars which at 7.30-7.35 Gyr are younger than 2 Gyr and at galactocentric radii $R<4\kpc$, and
     contribute to the high $\vlos$ peak ($210<\vlos/\kms<280$) at $l=6\pm1$ and $\left|b\right|<1$ degrees.
     We track them for 2.4 Gyr afterwards. We record the first, second and third time ($\tau_1$, $\tau_2$ and $\tau_3$) at which 
     they are at  $l=6\pm1$ and $210<\vlos/\kms<280$ and use it to calculate the first and second return time intervals $t_1=\tau_2-\tau_1$ and $t_2=\tau_3-\tau_2$.
{\it Panel A}: The histogram for $t_1$. The peak selection windows are marked by vertical lines in red, green, blue, purple and orange.
{\it Panels B-F}: $t_2$ histograms for orbits selected to be in one of the $t_1$ peak windows marked in Panel A. The colour of each panel
                  corresponds to the colour of the selection window. The selection windows from Panel A are marked by dotted vertical lines.
}
\label{hist}
\end{figure*}

These $\vlos$ distributions are shown in Figure \ref{vlos-peak}. Unsurprisingly they have their main peaks at the selection positions. 
Both distributions clearly show a distinct peak at negative $\vlos$ made up by orbits returning along the other side of the bar.
This peak was already visible in Figure \ref{vlos-age} for 9.3 Gyr. The increase in maximum absolute velocities
between 7.3 and 9.3 Gyr widens the gap between the peaks and shifts the main peak of the overall distributions for stars within $R<4\kpc$ to higher $\vlos$. Consequently the negative $\vlos$ peak
at 9.3 Gyr for Figure \ref{vlos-age} becomes visible. The negative peak comprises approaching stars while the positive peak features stars moving away from the sun.
Figure \ref{vlos-peak} reveals a third peak for 7.3 Gyr indicating that a group of orbits has a more complicated structure than alternately contributing to 
the two peaks.

In the right panel of Figure \ref{vlos-peak} we visualise the origin of the high and low $\vlos$ peaks at $230<\vlos/\kms<300$ and $-150<\vlos/\kms<-50$
for young bar stars at 9.3 Gyr from spatially  separate components along the line of sight. Here we plot the spatial positions of all stars in the 
positive $\vlos$ peak at 9.3 Gyr in black and of all stars in the negative $\vlos$ peak at 9.3 Gyr in red. The components clearly separate to the expected sides.
One orbit family that would contribute to both peaks are the $x_1$ orbits. In the rotating frame of the bar, these orbits move along opposing sides of the bar major
axis and turn around at the bar tips. They would obviously produce the two peak feature in the left panel of Figure \ref{vlos-peak}, as was e.g. discussed by \citet{molloy}. 
We overplot in green one such orbit extracted from the simulation which alternately contributes to both peaks. However, as we will show below, the high $\vlos$ features
in our simulation are made up of a more complicated population of orbits from different orbit families.

\begin{figure*}
\centering
\vspace{-0.cm}
\includegraphics[width=18.cm]{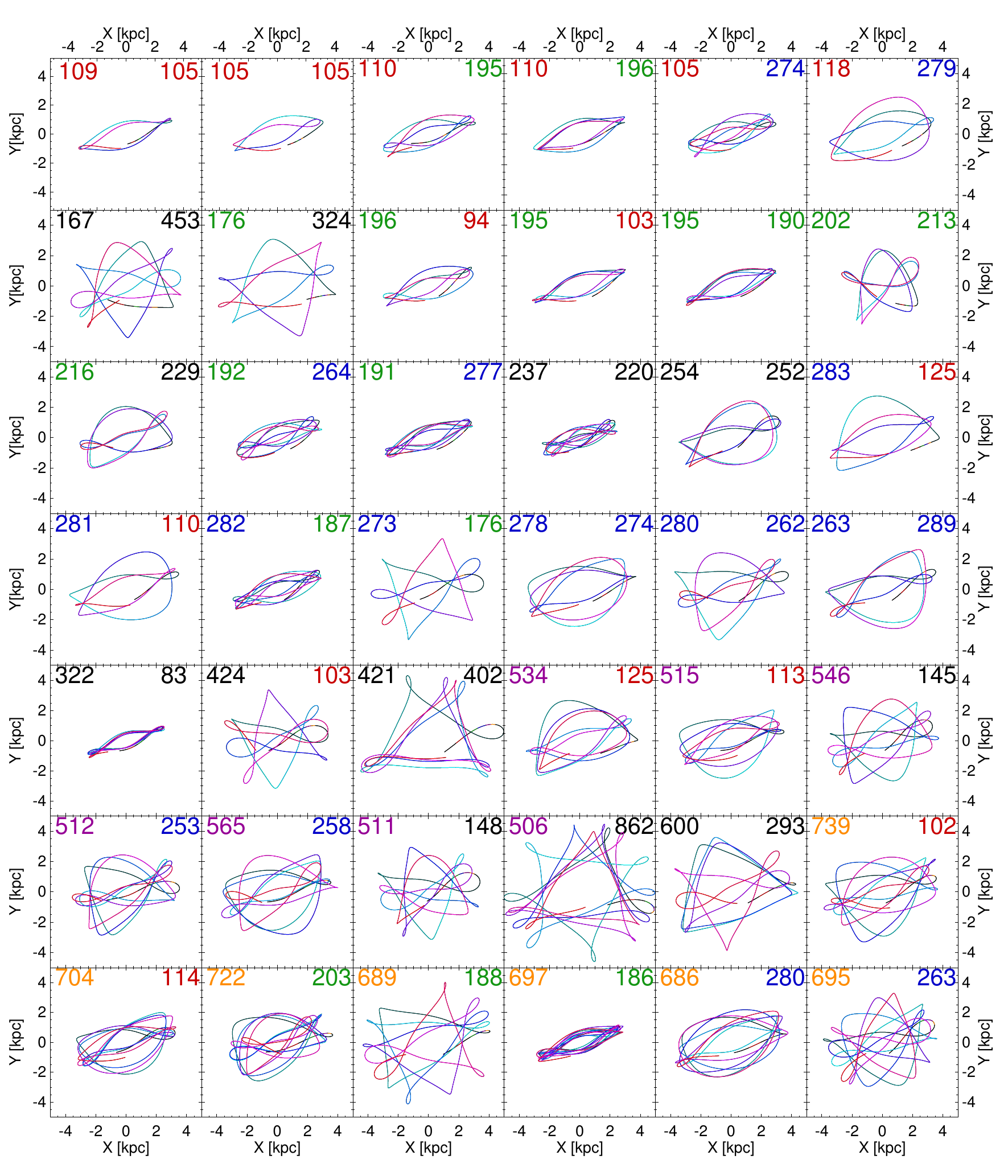}\\
\caption
{42 Orbits in the $x-y$ plane of particles that at 7.3 Gyr contribute to the high $\vlos$ peak at $l=6\pm1$ degrees. The orbits are shown from when they first leave the sightline until the second time
they return to the sightline and contribute to the high $\vlos$ peak again. The numbers in the corners of the panel indicate the times between contributions to the peak in Myr.
If these fall into one of the return time peaks identified in Figure \ref{hist} they are coloured accordingly, if they don't, they are coloured black.
The orbit colour coding indicates time, changing from black via light and dark blue to red with increasing time. 
}
\label{orbits}
\end{figure*}

\subsection{A simple classification for high $\vlos$ orbits}

To get a better understanding for the orbits that contribute to the peak, we track all orbits that contribute to the high $\vlos$ peak at $l=6\pm1$ and $\left|b\right|<1$ degrees
at 7.30-7.35 Gyr defined by $210<\vlos/\kms<280$ for young stars with ages $<2$ Gyr and radii $R<4\kpc$.
If an orbit is similar to a closed orbit, we will expect it to return to the same region in phase space periodically.
Our selection of sightline plus $\vlos$ range can be seen as a rough proxy for a phase space region. Considering that the bar grows and the potential changes, a more
sophisticated definition is complicated. Therefore we record the time interval needed by an orbit to return to the sightline and $\vlos$ window. The histogram for the
first return time $t_1$ is plotted in the Panel A of Figure \ref{hist}. 
We see that this distribution shows three distinct peaks at circa 110 (red), 190 (green) and 280 Myr (blue), which contain approx. 23, 13 and 28 per cent of the stars. 
There are also two broader peaks at 450-650 Myr (14 per cent, purple) and 650-800 Myr (7 per cent, orange).
Note that $\sim$ 13 per cent of orbits never return or have return times $>800$ Myr.
Were all these orbits simple $x_1$ orbits, we would expect a simpler structure.

In Panels B-F of Figure \ref{hist}, we look at histograms of the second return time $t_2$ for stars selected to have first return times $t_1$ in the five peaks found
in the $t_1$ histogram. To some degree, all these $t_2$ histogram resemble the $t_1$ histogram, as they reproduce one to three of the previously found three narrow peaks.
Above 450 Myr, the $t_2$ histograms all seem dominated by noise, but all of them contain a similar fraction of such orbits, varying between 25 and 31 per cent.
A first clear conclusion is, that for each histogram, the fraction of orbits that remain in the same return time peak is below 50 per cent.
This shows that most orbits do not show a steady periodicity in our definition, but rather combine a low number of possible return times.

To connect the shape of these histograms to types of orbits, we extracted random sets of orbits with similar combinations of return times $t_1$ and $t_2$, for example
contributors to the green peak for $t_1$ and to the blue peak for $t_2$. In addition we extracted a set of orbits that do not contribute to return time peaks.
From these we select by eye 42 typical orbits. They are shown in Figure \ref{orbits}. Note that due to different prominence of the peaks this figure does not 
represent orbit types at their underlying contribution fractions. 

The stars with shortest return times $t_1$ (red peak) show a high fraction (45 per cent) of stars with $t_1\approx t_2$. Additionally, 30 per cent of the stars have $t_2$ in the area 
of the green or blue peak, but these peaks are not separately visible. The first row of Figure \ref{orbits} reveals that these stars are essentially all on $x_1$ like orbits.
If they have $t_2\gg t_1$, it is mostly because the orbits are not exactly closed. They can miss the high $\vlos$ window as their directions of motion and velocity 
vary when they cross the sightline.  Moreover, $N$-body noise can cause the integrals of motion of the particles to fluctuate.

Stars from the second (green) $t_1$ peak show three $t_2$ peaks at the locations of the three narrow $t_1$ peaks which are equally populated ($\sim25$ per cent each).
Most stars are again on $x_1$-like orbits, but there
are also the first objects from more complex closed orbit families with Pretzel-like structures \citep{portail2}, such as the 202-213 or 216-229 objects.
At higher return times, $x_1$ orbits are a minority. If they have such return long return times, it is because they miss $\vlos$ due to the fluctuations mentioned above
more often than not, e.g. because they have lower energies and only in exceptional cases reach high enough $\vlos$.

Pretzel-like morphologies are the most frequent pattern found at longer return times. For blue $t_1$ stars, there is again a high fraction (46 per cent) of $t_1\approx t_2$
stars, which tend to shows such patterns, as seen in the fourth row of Figure \ref{orbits}. Generally, the structures become more complicated and more irregular
as $t_1$ increases. Clearly some orbits show 3:1 characteristics with a changing orientation (objects 421-402 or 506-862), but there are also more complicated,
but apparently regular structures, such as the object 273-176, which was labelled 'Q' in \citet{voglis}, and objects like 176-324 or 424-103 which also combine 3:1 
and $x_1$ features. Other orbits appear to change structure with time, an indication of chaotic behaviour.

If we combine the results from above, we estimate that the fraction of stars in the high $\vlos$ peak that are on $x_1$ orbits
at the given time is between 40 and 50 per cent.
If we determined the full population of stars, which over a longer timescale ($\sim1 $ Gyr) periodically contributed to the high $\vlos$ peak,
the fraction of $x_1$ stars would be even lower, as they have the shortest return times.
This invalidates the original assumption of \citet{molloy} that the high $\vlos$ feature is only caused by 2:1 orbits.
There are other major contributors, most prominently the Pretzel-like orbits, but also 3:1 or more complex orbits.
We note that \citet{molloy} have recently updated their paper in response to our work and now also find that higher order orbit
families (they name 3:1 and 5:2) can make significant contributions to the high $\vlos$ features.

We note that we have attempted to analyse the vertical structure of orbits in connection to the in-plane one and
experimented with frequency analysis, but concluded that the results were to diverse or noisy to help with the understanding
of the high $\vlos$ features.

\subsection{Are orbits evolving away from high $\vlos$ orbits?}

\begin{figure}
\centering
\vspace{-0.cm}
\includegraphics[width=8cm]{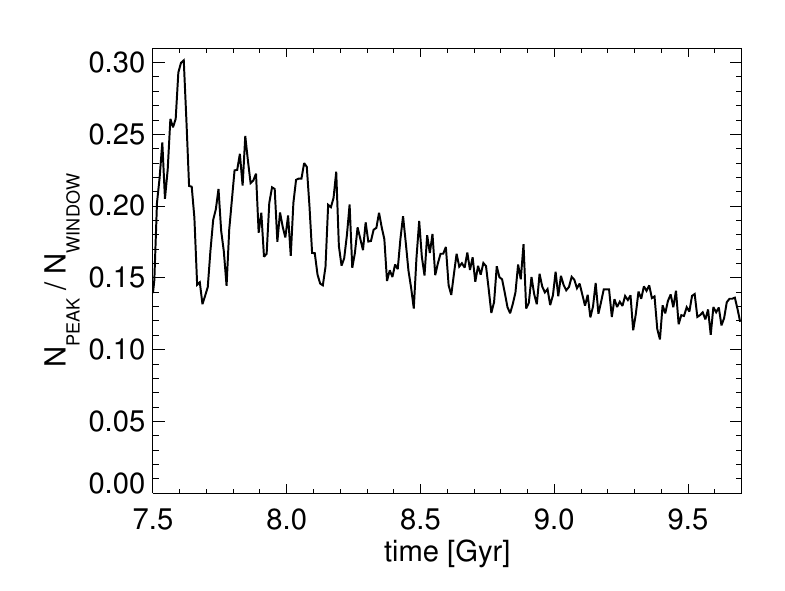}\\
\caption
    { For this Figure, we track all orbits from the histogram in Panel A of Figure \ref{hist}.
      We exclude orbits that never return to the high $\vlos$ window or have very long return times $t_1>$ 0.8 Gyr. 
      We record each passage of the sightline $l=6\pm1$ degrees (total number $N_{\rm window}(t)$) and each contribution 
      to the high $\vlos$ peak $210<\vlos/\kms<280$) during such a passage (total number $N_{\rm peak}(t)$). 
      We plot the fraction $\nu(t)=N_{\rm peak}(t)/N_{\rm window}(t)$.
}
\label{frac}
\end{figure}

In our discussion of Figure \ref{hist}, we already noted that around 10 per cent of stars which contribute to the high $\vlos$ peak at a given time and sightline,
need 1 Gyr or longer to achieve this again. The $t_2$ histograms have also shown that $\sim$25 per cent of stars that initially had return times $<400$ Myr
need more than 500 Myr to return again. So orbit populations slowly change in a way that reduces the proportion of their time during which they contribute
to the high $\vlos$ peak.

To understand why the high $\vlos$ peak vanishes with age, we conducted the following exercise: We tracked all orbits from the histogram in Panel A of Figure \ref{hist}.
As the orbits that never return or have very long return times (defined as $>$ 0.8 Gyr) are likely random contributors, we exclude them. We record each passage of the sightline
(total number $N_{\rm window}(t)$) and if during this passage the orbit contributes to the high $\vlos$ peak (total number $N_{\rm peak}(t)$). The fraction $\nu=N_{\rm peak}(t)/N_{\rm window}(t)$
gives a measure of how much the population of orbits is contributing to the peak at time $t$. 
A priori one would expect a ratio $\nu \sim 0.5$ for $x_1$ orbits as they are approximately axisymmetric along the bar passing once through the sightline while approaching
and once while moving away and contributing to the high $\vlos$ peak. Consequently, smaller values for $\nu$ are expected for more complex orbits.
This is slightly complicated by the geometry and the slower absolute velocity of stars while approaching: The front side of the bar is geometrically disfavoured, 
as the pencil beam covers a larger area at the rear side. On the other hand, the approaching stars are closer to the tip and hence slower, taking more weight in 
the observations. Moreover, it is also possible that stars moving along the rear side have high enough $\vlos$ only for a fraction of the time they spend in the sightline.
If we select $x_1$ orbits which at every second crossing of the sightline contribute to the high $\vlos$ peak, we empirically find a mean value of $\nu \approx 0.45$, 
close to the simple estimate.

\begin{figure*}
\centering
\vspace{-0.cm}
\includegraphics[width=13.5cm]{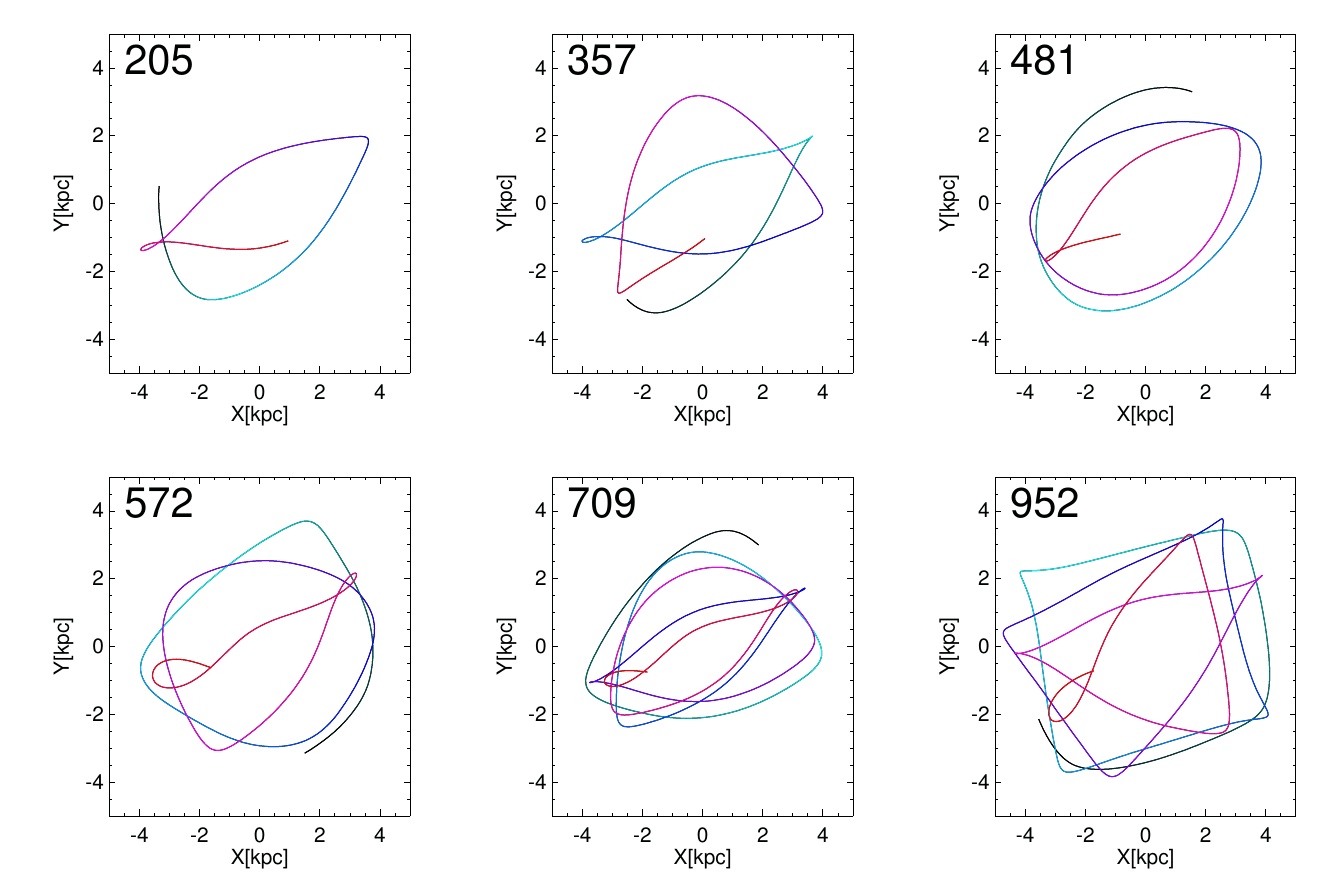}\\
\caption
    { Orbits of captured particles in the $x-y$ plane. These stellar particles are born between 7.3 and 8.3 Gyr and contribute to the high
      $\vlos>220\kms$ feature at $l=6\pm1$ degrees and 8.3 Gyr. The numbers in the top left corners indicate the time in Myr between birth and first detection 
      in the high $\vlos$ window at $l=6\pm1$ degrees. The colour coding indicates time, from black at birth via light and dark blue to red at first detection.
}
\label{capture}
\end{figure*}

From the return time analysis shown above, we would expect $\nu<0.45$, as most orbits exhibit a more complex behaviour.
$\nu(t)$ is shown in Figure \ref{frac}. Initially, $\nu$ fluctuates periodically. This can be understood as most orbits return to the high $\vlos$ peak after specific characteristic times
(see Figure \ref{hist}) and they have been selected at correlated phases. The average value of $\nu$ over the covered 2 Gyr drops from $\nu\sim 0.22$ to $\nu\sim0.12$. 
This shows that the fraction of orbits contributing to the high $\vlos$ peak continuously decreases. 

Most of this decrease is caused by diffusion of stars away from the contributing orbits, but it is not clear, if our simulation handles this diffusion correctly. Sources for this 
diffusion are chaotic orbit behaviour, two-body noise and fluctuations in bar parameters. Our resolution is likely not sufficient to cover chaotic orbital behaviour correctly. For the 
two-body noise, it is unknown if the level of $N$-body noise is realistic compared to the MW (bearing e.g. Giant Molecular Clouds near the bar tips, and dark matter substructure). 
In addition, our simulation will not have sources for bar fluctuations like tidal torques exerted by satellites. 

The important point to take away is that the vanishing of the high $\vlos$ peak in older stars is not only due
to the broadening of the main peak, but also due to active loss of stars from high $\vlos$ orbits.

\subsection{Orbit capture by the bar}

By design, new-born star particles in our simulation are released on near circular orbits, where the circular speed was determined to balance the local radial component
of the gravitational force. Obviously, the young stars contributing to the high $\vlos$ peak discussed above are not on circular orbits. Just as obviously, there are no stable
circular orbits near the bar. This is why we set an inner radius limit for new born stars, so that no new star particles are inserted within the main bar region.
This cutoff is, of course, to some degree arbitrary and young stars born close to the bar can be captured by the bar, resulting in the young bar populations studied in
the previous Sections. Bars in real galaxies can also grow by capturing stars and thus changing their orbits.
So the young stars that end up in the high $\vlos$ peak are predominantly the result of continuous bar growth.

In Figure \ref{capture}, we show six stars that at 8.3 Gyr contribute to the high $\vlos>220\kms$ feature and were born within 1 Gyr before that time. We find that
the shortest time it takes to appear in the high $\vlos$ peak is similar to the shortest return time peak, 110 Myr. The capture times are rather equally distributed
between this lower limit and the maximum 1 Gyr due to the selection criteria. The shown star particles were selected at different {\it capture time} intervals.
They start out on non-bar orbits and at some point transition to $x_1$-like or Pretzel-like orbits.

\begin{figure}
\centering
\vspace{-0.cm}
\includegraphics[width=8cm]{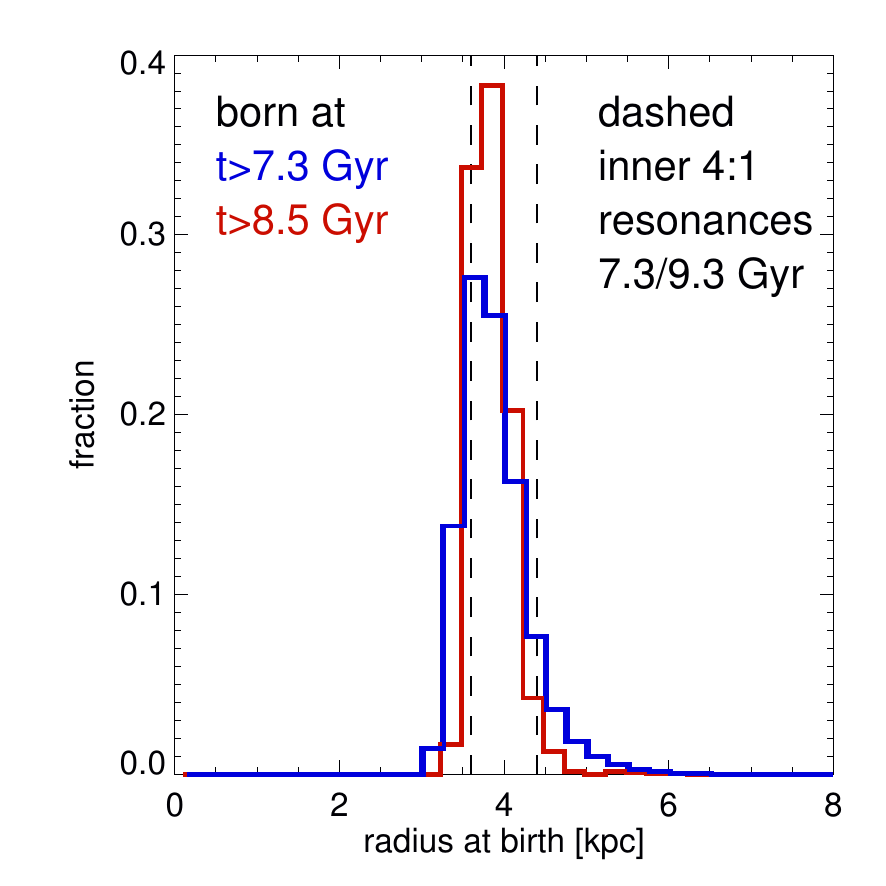}\\
\caption
    { For this Figure we select stars which were born after 7.3 Gyr and at 8.3-8.5 or at 9.3-9.5 Gyr contribute to the high $\vlos$ window at $l=6$ degrees.
      The union of the samples found at these two time intervals comprises $\sim11\,000$ stars. The distribution of birth radii $R_{\rm birth}$ of all $\sim11\,000$ stars
      is shown in blue. The histogram of stars born after 8.5 Gyr and contributing at 9.3-9.5 Gyr is shown in red. 
      The vertical dashed lines mark the positions of the inner 4:1 (ultra-harmonic) resonance at 7.3 and 9.3 Gyr.
}
\label{radbirth}
\end{figure}

Clearly stars that end up in the high $\vlos$ features are born at radii that correspond to the tip of the bar, 
so they belong to the innermost inserted particles. To visualise this point, we plot in
Figure \ref{radbirth} the distribution of birth radii $R_{\rm birth}$ of high $\vlos$ stars. For this Figure we select stars which were born after 7.3 Gyr and contribute to the
high $\vlos$ feature at $l=6$ degrees at 8.3-8.5 Gyr and/or at 9.3-9.5 Gyr. The union of the samples found at these two time intervals comprises $\sim11\,000$ stars,
for which we determine birth radii. The histogram for all stars is shown in blue, the one for the subsample of stars born after 8.5 Gyr is shown in red.
We notice that the majority of stars is born just outside the bar at $R\sim 4\kpc$. As seen in Figure \ref{bar-evo} the bar radius fluctuates mildly, which allows smaller
birth radii. The subsample of stars which were born at later times has its lower cutoff slightly further out due to the continuing growth of the bar.
The outer tail is wider for older stars, indicating that stars which are older at capture can have larger birth radii. This is because they had more time
to migrate radially inwards.

Resonances can play a role in capturing orbits (e.g. \citealp{bt}). The corotation resonance of our bar at the analysed time intervals is at $R=7-9\kpc$, but no stars
from these regions are captured. We find that the ultra-harmonic resonance (UHR), $4(\Omega-\Omega_{\rm bar})=\kappa$,
in our model lies just outside the bar. This is likely not coincidental, as the UHR can limit the radial extent of stable $x_1$ orbits and thus the size of the bar \citep{patsis}.
The resonance may play a role in shaping the spatial distribution from which stars are captured by the bar, 
as while the bar grows, the resonance moves outwards, and stars on orbits just outside the bar get scattered, absorbed or released by it.
These processes can transport orbits to the regions of phase space occupied by high velocity bar orbits. However, the exact 
role of the resonance in the capture process is unknown, as the bar is strong, evolving and the potential is not smooth.
The capture of orbits can be understood in simpler terms (similar to \citealp{lb}), in that
stars come in at the right angle to the bar to lose angular momentum and get trapped.
A detailed study of capture processes is beyond the scope of this work.

Capture of stars onto these orbits along the bar is mostly drawn from kinematically cool populations that are close to the Galactic plane, which have a small random 
energy compared to the potential disturbance and are at low altitude, where the disturbance is strongest. While these stars drastically change the in-plane geometry of 
their orbits during capture by the bar, their vertical energy and hence scale height remain small. Consistently, both the MW and our simulation display a 
discernible high $\vlos$ feature only at observed latitudes $b \lesssim 2$ degrees.

Above, we have learned that for populations of stars which at a given time contribute to the high $\vlos$ feature, the contribution fraction slowly
declines with time. As the simulation evolves, the high $\vlos$ feature, however, remains detectable. This is because of ongoing star formation close
to the bar region and the continuous capture of young stars on $x_1$-like or Pretzel-like orbits. The young populations on such orbits then slowly evolve
towards more complex orbit populations with a decreasing fraction of high $\vlos$ stars.

We note that another process which could produce young stars on high $\vlos$ orbits is not included in our model:
In real galaxies, molecular clouds can exist on bar orbits \citep{sheth02}. Stars born in these clouds would make up kinematically cool
populations born directly on high velocity bar orbits and could thus contribute to the high $\vlos$ features. 

\section{Conclusions}

In this paper, we have studied how an $N$-body+SPH model of a growing Milky Way (MW) like barred disc galaxy compares to the MW in terms of line of sight velocities $\vlos$
of stars seen from the earth and in the direction of the bar. The model uses an approach that has not yet been used in this context. We start with a light 
(one tenth of the final mass) and short scale-length disc of gas and stars in equilibrium within a live dark halo. This disc is grown by continuously adding star and gas
particles on near circular orbits in a controlled way. The mass and size growth histories are matched to expectations for the MW, and indeed at simulation time 7-10 Gyr
the simulated bar is similar in structure to the MW bar.

Our simple prescription for the evolution of the galaxy in our $N$-body simulation lowers the computational cost and also circumvents the uncertainties
in the parametrisation of sub-grid model approaches to star formation and stellar feedback. The general dynamics of gas and stars are expected to be reasonably represented,
but we do not cover varying gas fractions, supernova induced turbulence and effects of multiple gas phases. The simulation 
does not produce any sensible interplay between gas and stars: there is e.g. no enhanced star formation at places with high gas densities. External galaxies 
show indications for star formation associated with the gas flows along the edges of a bar \citep[the equivalent of the MW's $3 \kpc$ arms;][]{sheth02}, 
which depend strongly on the dynamical state of the bar \citep[see e.g.][]{friedli95}. However, these processes are not covered in our simulation. On the positive side, our approach allows 
us to set a reasonable star formation history, to keep spurious sources of heating under control, and hence to observe the dynamical effects of bar formation with less scatter.

We use this simulation to interpret line of sight velocity ($\vlos$) data for the central and inner disc regions of the MW from the APOGEE survey. 
To enable a robust comparison, we apply a simple population synthesis to model the APOGEE selection function. Apart from minor discrepancies at a
minority of sightlines, the model fits the data surprisingly well without any scaling or adjustment. The simulation not only reproduces the 
positions and dispersions of the main peaks in the $\vlos$ distributions at each sightline, but also displays the high $\vlos$ features found 
by \citet{nidever} within the central bar region. 

Our simulation clearly identifies these high $\vlos$ features with stars travelling near the plane alongside the bar. By showing that these are 
kinematically cool features with preferentially young stars in the plane, we naturally explain why the high $\vlos$ structure was seen by \citet{nidever} 
and \citet{babusiaux} in observation fields at $\left|b\right| \lesssim 2$, but is absent in the higher-latitude fields studied by \cite{zoccali}. By the 
strong preference of the APOGEE selection function for young stars, the high $\vlos$ features get additionally enhanced and have 
comparable strength and shape in our simulation and the APOGEE observations. We also determine at which velocities low $\vlos$ peaks at negative $l$, produced by stars moving
towards the Sun along the front side of the bar, should be visible. So far, only hints of these features have been seen by \citet{babusiaux}.

We show that, unlike for the high $\avlos$ features, the position and width of the main peaks of the $\vlos$ distributions are dominated by fore- and background stars,
which are not associated with the bar. We also study a variation of our parameters, in particular the bar angle $\phi$. Other than \cite{molloy} 
we feel unable to put strong constraints on the observed bar angle: while large $\phi \gtrsim 40 \deg$ are clearly disfavoured as the majority of high velocity 
stars get a too large projection angle with the line of sight, no reliable constraint can be found for smaller values of $\phi$, since 
change in velocities is degenerate with our general uncertainty about the bar length, pattern speed and strength. Additionally, we show
that the high $\vlos$ feature is detectable at simulation times between 7 and 10 Gyr and is thus not connected to specific events.
The feature changes mildly over this period due to a growth of the bar in length and mass, but the change is too small to allow 
any conclusions about the properties of the MW bar.

We note that on two sightlines, namely $l = 14$ and $l = 16$ degrees, the simulation shows a mild difference to the observations, missing the tentative 
high $\vlos$ peak at $l = 14$ and predicting the main peak of the velocity distribution at too high velocities at both sightlines. Possible origins
for the discrepancies are a longer bar length for the MW or the $4 \kpc$ molecular ring in the MW. A decisive answer on this question 
would require detailed analytical models that can cover the full space of possible parameters.

A detailed analysis of the stars participating in the high $\vlos$ features in our model reveals that \citep[contrary to the original view of][]{molloy}
only about half of the contribution is made by stars on $x_1$ orbits, with additional large contributions from Pretzel type \citep{portail2} 
and more complicated orbits. This finding results from a detailed study of these orbits guided by return times into the high velocity peak at 
the sightline $l = 6$ degrees. The return times display several peaks that are associated with the different orbit families (and some scatter by stars 
missing the observation window) and help with the quantification of the contribution of different orbit families. 
We note that, in response to our work, \citet{molloy} have updated their paper and now also conclude that higher order orbit
families (they name 3:1 and 5:2) can make significant contributions to the high $\vlos$ features.

Another interesting statistic is the ratio of the number stars in the high $\vlos$ region of a certain sightline divided by 
all stars passing through the sightline at arbitrary $\vlos$ values for all stars that have been 
observed once in the high velocity peak. While we find that $x_1$ orbits with stable short return times show an average ratio of 
$\nu \approx 0.45$, the stars in the simulation start with a ratio of $\nu \approx 0.22$, as a consequence of the significant contribution of more 
complex orbits to the velocity peak.

Within $2$ Gyr of the initial detection, this ratio drops steadily down to $\nu \approx 0.12$. Due to 
changes in the bar potential, length, and pattern speed, as well as orbital diffusion, stars diffuse out of these orbit families weakening 
the high $\vlos$ peak. However, the high $\vlos$ feature in our simulation persists, as the growing bar captures new stars from the surrounding disc. 
It is unclear, how realistically our model represents the precise balance of orbit capture and diffusion, which is determined by chaotic diffusion, two-body 
interactions and fluctuations in bar parameters. In addition, observations of enhanced star formation at the bar tips and on bar orbits along the leading bar edges (see above),
imply a source of young stars on high velocity bar orbits, which is not mirrored by our model. Overall, our result provides an indication for the recent growth 
of the Galactic bar. Quantitative constraints on the bar dynamics and history would demand further exploration of model parameters and also a detailed 
age and distance tomography of the observed high $\vlos$ stars.

We note that our bar is a slow bar with $R_{\rm bar}\sim 0.5 R_{\rm CR}$ in a disc with a high dark matter contribution to the circular velocity curve.
As discussed in Section 4, observations do not constrain the pattern speed and the length of the MW bar well and we currently can exclude neither a slow bar as ours
nor a fast bar with $R_{\rm bar}=0.7-1.0 R_{\rm CR}$. As far as the dark matter fraction is concerned, our disc agrees with recent estimates on the local circular 
velocity in the MW and the corresponding DM contribution. Further work is necessary to understand how faster bars and/or lower dark matter fractions would
alter the processes discussed in this paper.

\citet{deb15} point out that a nuclear stellar disc with a radial extent of $R\sim1-1.5\kpc$ could explain high velocity features at $l\sim10$ degrees.
Although this idea is interesting, they do not currently have a model which can explain the velocity distributions at the full APOGEE range of bar longitudes, as our model can.
The simulations of \citet{sormani} advocate against such a giant $x_2$ disc for the current MW centre, but the extent of such a disc may have varied over time.
To safely determine whether this idea can explain the data, a model, which self-consistently predicts velocity distributions at longitudes $l=(-20)-20$ degrees 
would be needed. A higher number of observed stars in this region with reasonable measurements of distances, ages and metallicities would certainly help
to (in)validate our model or other attempts to explain the observed velocity distributions. As far as the nuclear stellar disc in the MW is concerned, its existence so far
has only reliably been detected in the central $150\pc$ \citep{nudi} and the short scalelength inferred for this structure currently seems to disfavour a large nuclear 
disc as proposed by \citet{deb15}.

To summarise, we have shown that the high $\vlos$ features seen towards the bar in APOGEE are likely due to preferentially young stars on orbits moving
along the major axis of the bar. They move close to the plane, which is why the feature disappears at high latitudes. The most important 
orbit family contributing to the feature are $x_1$ orbits, but a variety of more complex orbits also make significant contributions.
Despite its simplicity, our model is the first shown to match the detailed kinematics in the bar region, and hence its late stages 
can serve as a blueprint for further and more detailed studies.

\section*{Acknowledgements}
It is a pleasure to thank James Binney for helpful discussions and comments on the manuscript. We are grateful to Mattia Sormani and Evgeny Vasilyev for helpful comments.
The simulations presented here were run on the DiRAC facility jointly funded by STFC, the Large Facilities Capital Fund of BIS and the University of Oxford.
This work was supported by the UK Science and Technology Facilities Council through grant ST/K00106X/1 and by the European Research Council under the European Union's
Seventh Framework Programme (FP7/2007-2013)/ERC grant agreement no.~321067.

\end{document}